\documentclass{aa}

\usepackage{graphicx}

\usepackage{subfig}
\usepackage{array}

\usepackage{txfonts}
\usepackage{natbib}
\usepackage{amssymb}
\usepackage{multirow}
\usepackage{adjustbox}
\bibpunct{(}{)}{;}{a}{}{,} 
\usepackage{pdflscape}
\usepackage{supertabular}
\usepackage{arydshln}
\usepackage{bigdelim}

\begin{document}

   \title{Infrared-Faint Radio Sources in the SERVS deep fields:}
   \subtitle{Pinpointing AGNs at high redshift.}

   \author{A. Maini \inst{1,2,3,4}
          \and I. Prandoni \inst{3}
          \and R. P. Norris \inst{4,5}
          \and L. R. Spitler \inst{2,6}
          \and \\A. Mignano \inst{3}
          \and M. Lacy \inst{7}
          \and R. Morganti \inst{8,9}
          }

   \institute{Dipartimento di Fisica e Astronomia, Universit\`a di Bologna, viale B. Pichat 6/2, 40127 Bologna, Italy
             \and Department of Physics and Astronomy, Macquarie University, Balaclava Road, North Ryde, NSW, 2109, Australia
             \and INAF-IRA, via P. Gobetti 101, 40129 Bologna, Italy
             \and CSIRO Astronomy \& Space Science, PO Box 76, Epping, NSW 1710, Australia
             \and Western Sydney University, Locked Bag 1797, Penrith South, NSW 1797, Australia
             \and Australian Astronomical Observatories, PO Box 915, North Ryde, NSW 1670, Australia
             \and NRAO, 520 Edgemont Road, Charlottesville, VA 22903, USA
             \and Netherlands Foundation for Research in Astronomy, PO Box 2, 7990 AA, Dwingeloo, The Netherlands
             \and Kapteyn Astronomical Institute, University of Groningen, PO Box 800, 9700 AV Groningen, the Netherlands
             }

   \date{Received ; accepted }

   \abstract
    {Infrared-Faint Radio Sources (IFRS) represent an unexpected class of objects relatively bright at radio wavelength, but unusually faint at infrared (IR) and  optical wavelengths. A recent and extensive campaign on the radio-brightest IFRSs (\mbox{$S_{1.4\,GHz}$ $\gtrsim$ 10\,mJy}) has provided evidence that most of them (if not all) contain an AGN. Still uncertain is the nature of the radio-faintest ones ($S_{1.4\,GHz}$ $\lesssim$ 1\,mJy).    
    }
    {The scope of this paper is to assess the nature of the radio-faintest IFRSs, testing their classification and improving the knowledge of their IR properties making use of the most sensitive IR survey available so far: the \textit{Spitzer} Extragalactic Representative Volume Survey (SERVS). We also explore how the criteria of IFRSs can be fine-tuned to pinpoint radio-loud AGNs at very high redshift ($z$ $>$ 4).
    }
    {We analysed a number of IFRS samples identified in SERVS fields, including a new sample (21 sources) extracted from the Lockman Hole. 3.6 and 4.5\,$\mu$m IR counterparts of the 64 sources located in the SERVS fields were searched for, and, when detected, their IR properties were studied.
    }
    {We compared the radio/IR properties of the IR-detected IFRSs with those expected for a number of known classes of objects. We found that they are mostly consistent with a mixture of high-redshift ($z$ $\gtrsim$ 3) radio-loud AGNs. The faintest ones (\mbox{$S_{1.4\,GHz}$ $\sim$ 100\,$\mu$Jy}), however, could be also associated with nearer ($z$ $\sim$ 2) dust-enshrouded star-burst galaxies. We also argue that, while IFRSs with radio-to-IR ratios $>$ 500 can very efficiently pinpoint radio-loud AGNs at redshift 2 $<$ $z$ $<$ 4, lower radio-to-IR ratios ($\sim$ 100--200) are expected for higher redshift radio-loud AGNs.   
    }
    {}


   \maketitle


\section{Introduction}
\label{sec:Intro}

Infrared-Faint Radio Sources (IFRS), first discovered by \citet{Nor2006}, are a serendipitous by-product of the routine work of cross-matching catalogues taken at different wavelengths. The first IFRSs were identified in the CDFS and in the ELAIS-S1 fields, cross-matching the deep \mbox{($S_{1.4\,GHz} \geq$ 50--100\,$\mu$Jy)} Australia Telescope Large Area radio Survey (ATLAS; \citealt{Nor2006}; \citealt{Mid2008a}), with the \textit{Spitzer} Wide-area IR Extragalactic Survey (SWIRE\footnote{http://swire.ipac.caltech.edu/swire/swire.html.}, \citealt{Lon2003}) at all the IRAC bands (3.6, 4.5, 5.8 and 8.0\,$\mu$m) and at the 24\,$\mu$m MIPS band.

These sources were identified as an interesting class of objects due to their lack of any IR counterpart, down to the SWIRE detection limit \citep{Nor2006}. Given the sensitivity of the SWIRE survey it was expected that extragalactic radio sources within $z$ $\sim$ 2 belonging to any known class of objects would be detected, regardless of whether the radio
is generated by star formation or AGN activity, for any reasonable dust obscuration and evolutionary model. IFRSs showed a radio-to-IR flux density ratio of the order of 100 or above.

IFRS searches were later extended to the \textit{Spitzer} extragalactic First Look Survey (xFLS; \citealt{Gar2008}) using both the 610 MHz Giant Microwave Radio Telescope (GMRT) and the 1.4\,GHz Very Large Array (VLA), and to the SWIRE \mbox{ELAIS-N1} field \citep{Ban2011} using the Dominion Radio Astrophysical Observatory (DRAO) 1.4\,GHz survey \citep{Tay2007, Gra2010, Ban2011}, for a total of 83 IFRSs catalogued. These early samples of radio sources lacking any \textit{Spitzer} counterparts cover a wide range of 1.4\,GHz radio fluxes (from tenths to tens of mJy, with a preference around 1--2\,mJy) and were collectively named `first generation' IFRSs \citep{Col2014}.

A first attempt to quantify the average IR flux of these objects was performed by \citet{Nor2006}, who did a stacking experiment of \textit{Spitzer} data in the CDFS field, at all the IRAC bands and at the 24\,$\mu$m MIPS band. Nothing was detected, implying a mean IR flux density for these objects well below the SWIRE sensitivities. \citet{Nor2006} obtained IR upper limits stacking either the full sample of 22 sources (e.g., $\lesssim$ 0.5\,$\mu$Jy at 3.6\,$\mu$m, the most sensitive IRAC band, and $\lesssim$ 0.03\,mJy at 24\,$\mu$m), or only the 8 radio brightest sources to avoid any possible contamination from artifacts (e.g., $\lesssim$ 0.8\,$\mu$Jy at 3.6\,$\mu$m, and $\lesssim$ 0.05\,mJy at 24\,$\mu$m). These values excluded the possibility that 1st generation IFRSs could simply belong to the dim tail of the distribution of usual classes of objects, at the same time implying that they have unusual IR properties. The result at 24\,$\mu$m in particular, shows that IFRSs strongly depart from the classical Mid-IR-radio correlation for star-forming galaxies (\mbox{$q_{24} = \log{(S_{24\,\mu m}/S_{1.4\,GHz})} = 0.84\,\pm\,0.28$}; \citealt{App2004, Boy2007, Mao2011}). This suggested from the very beginning a possible AGN-driven radio emission (\citealt{Nor2006, Gar2008, Zin2011}). The extreme ratios $R_{3.6}$ = $S_{1.4\,GHz}/S_{3.6\,\mu m}$ (typically \mbox{$\gtrsim$ 100}) suggested to \citet{Huy2010} and \citet{Mid2011} a possible link with another class of extreme objects, the high-redshift radio galaxies (HzRG). HzRG is a class of very rare sources selected from all-sky radio surveys on the basis of extreme radio-to-Mid-IR ratios ($R_{3.6}$ $\geq$ 200) and steep radio spectra ($\alpha \lesssim -1.0$\footnote{We define the spectral index following the convention: $S = \nu^{\alpha}$.}), and are typically identified with radio galaxies harbouring very powerful and obscured AGNs, at redshift $z$ $>$ 1 \citep{Sey2007}.

Despite these works, a comprehensive analysis of the nature of 1st generation IFRSs was challenging due to a wide variety of study-specific selection criteria. For instance, requiring the lack of an IR counterpart implies samples of IFRSs strongly dependent on the sensitivity of the IR survey under consideration, while fainter radio sources with no counterparts tend to have smaller $R_{3.6}$. The result was a very heterogeneous class of objects.

To overcome this limitation, \citet{Zin2011} developed a set of survey-independent criteria:

\begin{itemize}
  \item $R_{3.6}$ $>$ 500
  \item $S_{3.6\,\mu m}$ $<$ 30\,$\mu$Jy
\end{itemize}

The first of these quantifies the ratios between radio and IR flux densities, while the second requires that these objects are at cosmological distance (and is satisfied by the whole sample of 1st generation IFRSs, due to the lack of any IR counterpart at the SWIRE detection limit). It is noteworthy that Zinn criteria tend to exclude radio-faint IFRSs, as the $R_{3.6}$ criterion is more difficult to satisfy for them than the simple lack of IR counterpart. Applying these criteria, \citet{Zin2011} extracted a list of 55 sources from four fields (xFLS, CDFS, ELAIS-S1 and COSMOS), which represents the first sample of `second-generation' IFRSs.

Following these criteria and explicitly requiring a detected IR counterpart \citet{Col2014} compiled a list of 1317 IFRSs by cross-matching the Wide-Field Infrared Survey Explorer (WISE; \citealt{Wri2010}) catalogue with the Unified Radio Catalog (URC; \citealt{Kim2008}). Among these sources only 19 objects have spectroscopic information available, and are all identified as broad-line Type\,1 quasars in the range \mbox{2 $\lesssim z \lesssim$ 3}.

\citet{Her2014} targeted with the Very Large Telescope (VLT) four 2nd generation IFRSs in the CDFS, selected to have an $R$-band counterpart. For three of them they successfully measured spectroscopic redshifts in the range \mbox{2\,$\lesssim\,z\lesssim$\,3}, and due to the presence of broad emission lines of a few thousand km\,s$^{-1}$, they classified them as Type\,1 AGNs.

\citet{Her2015a} used the Very Long Baseline Array (VLBA) to target 57 IFRSs belonging to the \citeauthor{Col2014} list, and successfully detected compact cores in 35 of them. These targets were selected based on their proximity (distance \mbox{$<$ 1\,deg}) to a VLBA calibrator, and on their visibility during available filler time of the VLBA array. They span a large radio flux density range ($\sim$ 11 $\rightarrow$ 183\,mJy), which can be considered as representative of the full \citeauthor{Col2014} sample ($\sim$ 8 $\rightarrow$ 793\,mJy). This confirmed that compact cores lie in the majority (if not all) of 2nd generation IFRSs, establishing them as a class of radio-loud AGN. In a more recent paper, Herzog et al.\ (in press) performed a comprehensive analysis of the radio spectral energy distribution (SED) of 34 out of 55 sources belonging to the original \citet{Zin2011} list. The majority (85\%) of their subsample shows a steep radio SED ($\alpha$ $<$ $-$0.8), and a significant percentage (12\%) an ultra steep SED ($\alpha$ $<$ $-$1.3, typically associated with high-z radio galaxies, see e.g.\ \citealt{Mil2008}). Moreover, they found that some of these sources have a SED consistent with GHz peaked-spectrum (GPS) and compact steep-spectrum (CSS) sources, implying that at least some IFRSs are AGNs in the earliest stages of their evolution to FR\,I/FR\,II radio galaxies.

Since \citeauthor{Col2014} and \citeauthor{Her2014} samples were limited to relatively bright radio sources ($S_{1.4\,GHz}$ $>$ 9\,mJy), the nature of radio-faint IFRSs is still unclear. Many of the 1st generation IFRSs lie in the \mbox{sub-mJy} radio domain, a regime where radio sources can be associated with star-forming galaxies (SFG) (see, e.g., \citealt{Pra2001, Mig2008, Sey2008, Smo2015}).

In this paper we re-analyse the IFRS samples from \citet{Nor2006, Mid2008a, Ban2011}, improving the old analyses by exploiting the deeper \textit{Spitzer} Extragalactic Representative Volume Survey (SERVS\footnote{http://irsa.ipac.caltech.edu/data/SPITZER/SERVS}; \citealt{Mau2012}). SERVS is a deeper \textit{Spitzer} follow-up of the five SWIRE fields (CDFS, ELAIS-S1, ELAIS-N1, XMM-LSS, LH) undertaken as part of the Warm Mission. As a result, SERVS reaches 5$\sigma$ sensitivities of 1.9 and 2.2\,$\mu$Jy for the 3.6 and 4.5\,$\mu$m IRAC bands, respectively \citep{Mau2012}. Moreover, we perform a more comprehensive analysis of the IFRS population for several ranges of $R_{3.6}$, and we discuss how the $R_{3.6}$ criterion can be fine-tuned to better trace this population up to high redshift \mbox{($z$ $>$ 4)}.

After a short description of the available IFRS samples in the SERVS fields (Sect.\ \ref{sec:Samples}), we introduce a new sample extracted from the Lockman Hole (LH) field (Sect.\ \ref{sec:New}). Using recent 3.6 and 4.5\,$\mu$m images we search for IR counterparts of all these IFRSs: the deeper SERVS data allow us to increase the number of IR detections of radio-faint IFRSs and in several cases to get information in two IR bands (Sect.\ \ref{sec:Analysis}). We present a comparison of the radio/IR properties of these sources with those of several prototypical classes of objects (Sect.\ \ref{sec:Models}), and we discuss the possible nature and redshift distribution of these IFRS samples (Sect.\ \ref{sec:Radio/IR_prop}). Finally, in Sect. \ref{sec:Conclusion} we summarise our results.

Throughout this paper, we adopt a standard flat $\Lambda$CDM cosmology with H$_0$ = 70\,km\,s$^{-1}$\,Mpc$^{-1}$ and $\Omega_M$ = 0.30.


\section{Existing first-generation IFRS samples}
\label{sec:Samples}

\begin{table}[t!]
\caption{\label{tab:Fields} Main parameters of the radio, SWIRE and SERVS surveys, together with the number of 1st generation IFRSs identified in each SWIRE and SERVS 3.6\,$\mu$m field.
}
\centering
\renewcommand{\arraystretch}{1.2}
\addtolength{\tabcolsep}{-1pt}
\begin{tabular}{c c c c c c}
\hline\hline
         & 1.4\,GHz$^{(a)}$  & \multicolumn{2}{c}{SWIRE$^{(b)}$} & \multicolumn{2}{c}{SERVS$^{(c)}$} \\
Field    &       Area        &      Area       &      IFRSs      &       Area      &     IFRSs       \\
Name     &     (deg$^2$)     &    (deg$^2$)    &      (\#)       &     (deg$^2$)   &     (\#)        \\
 (1)     &        (2)        &       (3)       &      (4)        &        (5)      &      (6)        \\
\hline                                                                                               
CDFS     &        3.7        &       6.58      &       22        &        4.5      &       21        \\
ELAIS-S1 &        3.9        &      14.26      &       29        &        3.0      &       17        \\
ELAIS-N1 &       15.2        &       8.70      &       18        &        2.0      &       5         \\
\hline                                                                                               
Total    &                   &                 &       69        &                 &       43        \\
\hline                                                                                               
         &                   &                 &                 &                 &                 \\
\end{tabular}
\begin{minipage}{0.46\textwidth}
\small{$^{(a)}$ The 1.4\,GHz flux density limit is $\sim$ 100\,$\mu$Jy in CDFS and ELAIS-S1 fields, and is $\sim$ 275\,$\mu$Jy in ELAIS-N1 .}\\
\small{$^{(b)}$ The SWIRE coverage refers to the IRAC instrument only. The average SWIRE 5$\sigma$ flux density limits are 3.7, 5.4, 48 and 37.8\,$\mu$Jy in the four IRAC bands, respectively; and 0.23, 18, 150\,mJy in the three MIPS bands, respectively.}\\
\small{$^{(c)}$ the average SERVS 5$\sigma$ flux density limits are 1.9\,$\mu$Jy at 3.6\,$\mu$m band, and 2.2\,$\mu$Jy at 4.5\,$\mu$m band.}
\end{minipage}
\end{table}

Radio sources without any IR counterparts down to the detection limits of the SWIRE survey were found in three SWIRE fields (CDFS, ELAIS-S1, and ELAIS-N1), and were named 1st generation IFRS by \citet{Col2014}. These samples belong to different radio surveys, characterized by  different sensitivities.

CDFS and ELAIS-S1 fields were observed with the Australia Telescope Compact Array (ATCA) at 1.4\,GHz in Data Release 1 (DR1) of the ATLAS survey (\citealt{Nor2006} and \citealt{Mid2008a}). ATLAS covers about 3.5\,deg$^2$ in each field, down to a typical rms sensitivity of 20--40\,$\mu$Jy, with spatial resolutions of 11\arcsec $\times$ 5\arcsec and 10\arcsec $\times$ 7\arcsec, respectively. A new version (DR3) of the 1.4\,GHz ATLAS catalogues has been recently released \citep{Fra2015}, but we decided to keep using DR1 for consistency with previous works.

The ELAIS-N1 field was observed at 1.4\,GHz down to a rms sensitivity of 55\,$\mu$Jy (\citealt{Tay2007, Gra2010}). The observations were carried out with the Dominion Radio Astrophysical Observatory Synthesis Telescope (DRAO ST; \citealt{Lan2000}), and combined with higher resolution VLA follow-up data (3.9\arcsec $\times$ 3.9\arcsec) and the FIRST survey (\citealt{Whi1997}, 5\arcsec $\times$ 5\arcsec) to provide better positions \citep{Ban2011}.

The lack of IR counterparts down to the SWIRE detection limit biased these 1st-generation samples towards radio-fainter sources, resulting in samples characterised by smaller $R_{3.6}$ ratios ($\gtrsim$ 100). On the other hand, the lack of an IR counterpart is a tighter criterion than the $S_{3.6\,\mu m}$ $<$ 30\,$\mu$Jy one. As summarised in \mbox{Table \ref{tab:Fields}} (Column 4), 22 and 29 IFRSs were identified in the CDFS and ELAIS-S1 fields, respectively \citep{Nor2006, Mid2008a}. Another 18 IFRSs were identified in the ELAIS-N1 \citep{Ban2011}.

When searching for IR counterparts in the deeper SERVS fields two problems arise. First, SERVS mosaics cover smaller regions than the corresponding SWIRE ones. Therefore, the analysis is necessarily limited to sub-sets of 1st generation IFRSs, i.e.\ those located within the 3.6 and/or 4.5\,$\mu$m SERVS fields. Moreover, SERVS approaches the confusion limited flux density regime and some IFRSs lie in extremely crowded regions, increasing the likelihood of false cross-identifications. This forced us to reject three additional sources (one in the CDFS field and two in the ELAIS-N1 field: CS0283, DRAO6, and DRAO10). Of the original 69 1st generation IFRSs belonging to the three aforementioned SWIRE fields, we retain in our study only 43 sources located in the SERVS 3.6\,$\mu$m mosaics (see \mbox{Table \ref{tab:Fields}}, Column 6). We notice that the 4.5\,$\mu$m IRAC detector has a slightly shifted field of view with respect to the 3.6\,$\mu$m one (\citealt{Sur2005}). So the total number of 1st generation IFRSs within the footprint of the 4.5\,$\mu$m mosaics is slightly smaller (40 instead of 43, see Table \ref{tab:IFRS} for details).

\section{New sample in the LH SERVS field}
\label{sec:New}

The LH field was observed at 1.4\,GHz with the Westerbork Synthesis Radio Telescope (WSRT; Prandoni et al., submitted), covering about 6\,deg$^2$ with a spatial resolution of 11\arcsec $\times$ 9\arcsec. This survey produced a catalogue of about 6000 radio sources down to a 5$\sigma$ flux limit of 55\,$\mu$Jy. We searched for IFRSs in the LH field by cross-matching (searching radius of $\sim$ 2\,arcsec) this catalogue with the SERVS IR (3.6\,$\mu$m) images and catalogues, which cover about 4.0\,deg$^2$ in the region. The search for IFRSs in this field was performed following a set of {\it ad hoc} criteria, designed to include IFRSs as faint as for 1st generation IFRSs, at the same time minimising the risk of contamination from normal galaxies populations, typically characterized by $R_{3.6}$ values $\lesssim$ 100 (see \citealt{Nor2006}).

We therefore applied a looser threshold on the radio-to-IR flux ratio than the one applied by \citet{Zin2011}, which biases IFRS samples against the faintest radio sources, by retaining as IFRSs those radio sources with \mbox{$R_{3.6}$ = $S_{1.4\,GHz}/S_{3.6\,\mu m} \geq$ 200} when they have a counterpart in the SERVS catalogue, and $S_{1.4\,GHz}/(2.0\,\mu$Jy) $\geq$ 200 when they have no counterpart in the SERVS catalogue (assuming a value of 2.0\,$\mu$Jy as a representative SERVS 3.6\,$\mu$m detection limit). Moreover we included only IR counterparts unresolved at the scale of the SERVS Point Spread Function (PSF = 1.9\arcsec; see Sect.\ \ref{subsec:Cutouts} for more details), which roughly correspond to intrinsic sizes $\lesssim$ 32\,kpc at \mbox{$z$ $\gtrsim$ 1}. This size criterion excludes local (extended) sources that may be intrinsically faint, together with partially overlapping interacting galaxies, for which aperture photometry flux density measurements would be unreliable. This criterion is tighter than the $S_{3.6\,\mu m}$ $<$ 30\,$\mu$Jy constraint applied by \citet{Zin2011}, as none of our sources has $S_{3.6\,\mu m}$ greater than 10\,$\mu$Jy (see Table \ref{tab:IFRS} for details).

As a final step we performed a visual inspection of the 3.6\,$\mu$m images for all the candidate IFRSs, and we removed the sources in crowded regions (i.e.\ for which the cross-matching is doubtful), those of uncertain radio position (e.g.\ due to a complex radio shape), and those coinciding with catalogued radio lobes. In the end, in the LH field we retained 21 sources at 3.6\,$\mu$m, 19 of which fall inside the 4.5\,$\mu$m mosaic footprint.

In summary, our analysis was carried out on a total of 64 sources, 59 of which fall inside the 4.5\,$\mu$m mosaic footprint (see Table \ref{tab:IFRS} for details).


\section{Flux density measurement at 3.6 and 4.5\,$\mu$m}
\label{sec:Analysis}

Flux densities at 3.6 and 4.5\,$\mu$m for all the sources in the SERVS deep fields were measured on the SERVS mosaics using a standardised procedure. Our flux density measurement includes three steps: 1) Correction for systematic offsets between radio and IR catalogues; 2) Extraction of the IR image cutouts centred on the corrected radio positions; 3) Flux density measurement through the aperture photometry technique.

\subsection{Correction for radio-IR positional offsets and extraction of image cutouts}
\label{subsec:Offset}

Positional offsets were established by cross-matching the ATLAS and SERVS catalogues for the CDFS and ELAIS-S1 fields, and the WSRT and SERVS catalogues for the LH field. For the ELAIS-N1 field we used the FIRST survey (see Sect.\ \ref{sec:Samples}), as the original DRAO observations have too poor spatial resolution (42\arcsec $\times$ 69\arcsec).

For each field and SERVS frequency we cross-matched the radio and SERVS source catalogues, and derived the distribution of the separation between closest matches. We defined the systematic offsets as the position of the peak of the Gaussians fitting the radio-IR separation distributions. We show the result in \mbox{Table \ref{tab:Offsets}}. Some offsets are larger than the SERVS positional accuracies (\mbox{0.1--0.2\arcsec}), but all are well below the SERVS PSF (\mbox{$\sim$ 2\arcsec $\times$ 2\arcsec}; \citealt{Mau2012}).

We then corrected the radio positions for the corresponding offsets, and extracted 3.6 and 4.5\,$\mu$m cutouts from the SERVS mosaics, each centred at the corrected IFRS radio position. The cutouts are \mbox{99 $\times$ 99} pixels (i.e.\ 30 $\times$ 30 SERVS PSFs) wide.

\begin{table}[!t]
\caption{\label{tab:Offsets} Radio-IR positional offsets for each field, both for 3.6 and 4.5\,$\mu$m catalogues.}
\centering
\renewcommand{\arraystretch}{1.2}
\begin{tabular}{c c c c c}
\hline\hline
Field    & \multicolumn{2}{c}{3.6 $\mu$m band} & \multicolumn{2}{c}{4.5 $\mu$m band} \\
Name     &    $\Delta$RA    &    $\Delta$Dec   &    $\Delta$RA    &    $\Delta$Dec   \\
         &     (arcsec)     &     (arcsec)     &     (arcsec)     &     (arcsec)     \\
\hline
CDFS     &       0.03       &    $-$0.04       &    $-$0.02       &    $-$0.20       \\
ELAIS-S1 &    $-$0.05       &    $-$0.02       &    $-$0.19       &    $-$0.11       \\
ELAIS-N1 &       0.04       &    $-$0.01       &    $-$0.19       &    $-$0.14       \\
LH       &    $-$0.01       &    $-$0.14       &       0.03       &       0.09       \\
\hline
\end{tabular}
\end{table}

\subsection{Aperture photometry at 3.6 and 4.5\,$\mu$m}
\label{subsec:Cutouts}

For each cutout, we used the \verb+DAOPHOT+ IRAF\footnote{http://iraf.noao.edu/} task to measure the flux density through the aperture photometry technique, either at the centre of the cutouts (when no source was visually identified) or at the nearest IR source position (when a likely IR counterpart was present). We set the aperture radius to 1.9\arcsec, which roughly corresponds to the PSF of SERVS point sources \citep{Mau2012}. Finally, we multiplied the measured flux densities for the IRAC aperture corrections corresponding to that radius (1.359 for the 3.6\,$\mu$m band and 1.397 for the 4.5\,$\mu$m band; \citealt{Sur2005}).

The uncertainties in the measured aperture flux densities were again estimated using the \verb+DAOPHOT+ task. We built a grid of 15 $\times$ 15 apertures for each cutout, and we computed the flux density from each aperture. For each cutout, the distribution of such flux densities was fitted by a Gaussian. Then, we iteratively rejected the $>$3$\sigma$ outliers (likely contaminated by the presence of a source) until the process converged. The standard deviation of the final Gaussian distribution is a measure of the local rms of the cutout and provides an estimate of the error on the flux density measurement. In case of non-detections a (3$\times$ local rms) upper limit is provided.

This procedure allowed us to look for detections below the 5$\sigma$ threshold of the SERVS catalogues. We notice that some of the brightest sources are listed in the SERVS catalogues, and the flux densities we derived are consistent with those provided in the catalogues. On the other hand our flux density errors are larger, and have to be considered as more conservative. For consistency we adopted our flux density (and error) estimates for all sources, disregarding any presence in the SERVS catalogues.

\begin{figure*}[t!]
\centering
\includegraphics[width=180mm]{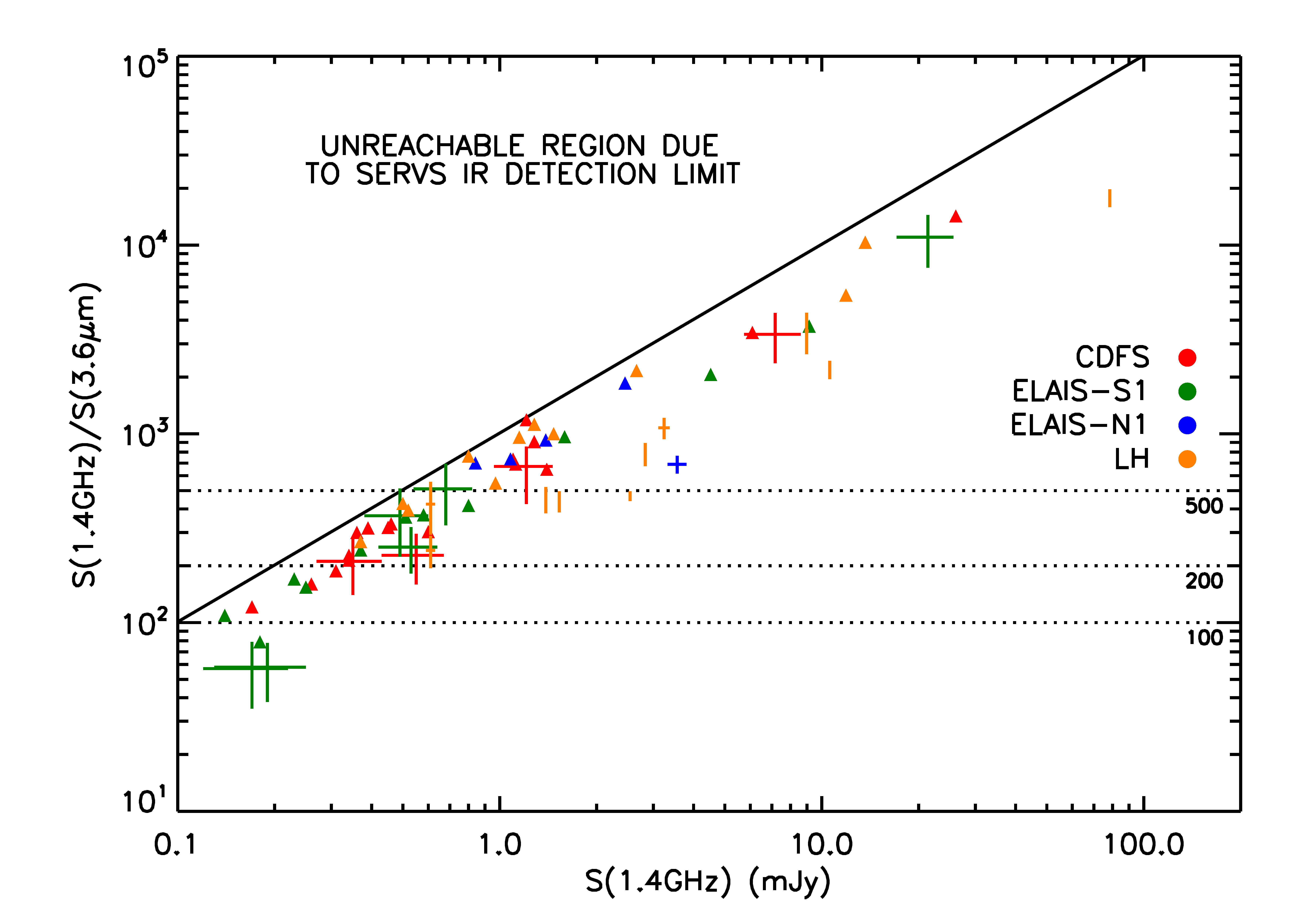}
\caption{\label{fig:Flux_Plotter} 1.4\,GHz-to-3.6\,$\mu$m flux density ratios versus 1.4\,GHz radio flux density. The diagonal solid line marks the lowest $R_{3.6}$ values we can trace due to the combined radio and SERVS detection limits (derived assuming \mbox{$S_{1.4\,GHz}$ $>$ 100\,$\mu$Jy} and \mbox{$S_{3.6\,\mu m}$ $>$ 1\,$\mu$Jy}, i.e. the smallest lower limit we measure in our sample). The dotted lines indicate flux density ratios $R_{3.6}$ = 100, 200 and 500. Flux ratios for reliably IR-detected IFRSs are reported as crosses (position with errors bars), while flux ratios of IR undetected/unreliable sources are indicated by triangles (lower limits). To each field corresponds a different colour, as explained in the legend.
}
\end{figure*}

\subsection{Reliability of SERVS counterparts}
\label{subsec:Cross-match}

We used the likelihood ratio technique (e.g.\ \citealt{Cil2003}) to compute the reliability of the SERVS counterparts, i.e.\ the probability of finding neither a chance identification nor a peak noise closer to the radio source than the IR candidate. The reliability was computed for both 3.6 and 4.5\,$\mu$m counterparts, and typically it ranges from 70\% to 99\% (Column 9, Table \ref{tab:IFRS}). To avoid contamination by false-positive identifications, in any following analysis we only retain SERVS counterparts with reliability \mbox{$>$ 90\%}, while we consider as undetected all IFRSs with no (or unreliable) SERVS counterparts.

In total we found 21 reliable counterparts at 3.6 and 20 at 4.5\,$\mu$m, for a total of 25 distinct IFRSs out of the original 64 sources we analysed (see Sect.\ \ref{sec:New}). When identified at both bands the reliability constraint ($>$90\%) is satisfied at both 3.6 and 4.5\,$\mu$m. The SERVS cutouts of all IFRS with reliable SERVS counterparts are shown in Appendix \ref{App:IFRS_CP}. For the LH IFRSs we also show the radio contours.

We also estimated the false detection rate, by searching for serendipitous detections with reliability \mbox{$>$ 90\%}. This was done by shifting the positions of the radio sources by an amount between 50\arcsec and 500\arcsec in 100 steps spiralling outward, and measuring the aperture photometry at each of such positions. False detection rates range from 5 to 8\% (depending on the field and on the band), so we expect that around 2 counterparts out of the 41 (21 at 3.6 and 20 at 4.5\,$\mu$m) retained as reliable IFRS are false.

\subsection{Information at other wavelengths}
\label{subsec:Other_CP}

The only previous attempt to search for IFRS IR counterparts using SERVS data was performed by \citet{Nor2011} who used a pre-release of the 3.6\,$\mu$m images. They detected 3 out of 39 1st generation IFRSs in the CDFS and ELAIS-S1 SERVS fields (CS0114, with \mbox{$S_{3.6\,\mu m}$ = 2.20\,$\pm$\,0.54\,$\mu$Jy}; CS0173, with \mbox{$S_{3.6\,\mu m}$ = 2.14\,$\pm$\,0.65\,$\mu$Jy}; and CS0255, with \mbox{$S_{3.6\,\mu m}$ = 1.91\,$\pm$\,0.53\,$\mu$Jy}), but concluded all detections were consistent with being chance associations caused by confusion. These three sources are detected also by us, with similar flux densities. Because of our slightly deeper SERVS mosaics and improved cross-matching algorithm, we are confident that most of our detections are not chance associations.

In Table \ref{tab:IFRS} we list the 3.6 and 4.5\,$\mu$m aperture flux densities and errors (or the 3$\sigma$ upper limits in case of no detection or unreliable SERVS identification) measured for all the sources in the SERVS fields (Columns 11 and 12). Also listed are the derived radio-to-infrared flux ratios or upper limits (Columns 13 and 14), and the 4.5-to-3.6\,$\mu$m flux ratios or upper/lower limits in case of detection at only one band (Column 15). The new SERVS mosaics allowed us to get reliable IR counterparts for 21 IFRSs at 3.6\,$\mu$m, for 20 at 4.5\,$\mu$m, and for 16 in both bands. All detected sources have IR fluxes of a few $\mu$Jy, typically with S/N $\sim$ 3--10. The three sources associated with crowded regions and discarded (see Section \ref{sec:Samples}) are marked with `--' in all IR-related columns. The IFRSs which are located outside the 4.5\,$\mu$m footprint are marked `out' in the 4.5\,$\mu$m flux density column.

For six sources we have found an optical counterpart. These counterparts have been identified by imposing a maximum search radius of 2\arcsec around the radio source position, and belong to a number of independent surveys as listed at the end of \mbox{Table \ref{tab:IFRS}}. With only one exception (LH3817, with \mbox{$K_{AB}$ = 20.75}), the optical magnitudes of these objects are very faint (AB magnitudes $\gtrsim$ 24, see Table \ref{tab:IFRS}, Col.\ 17), suggesting high redshifts. None of the reliable optical counterparts has a measured redshift.

\subsection{IFRS radio-to-IR ratio distribution}
\label{subsec:ratios}

Figure \ref{fig:Flux_Plotter} shows the $R_{3.6}$ ratios (or corresponding lower limits) versus 1.4\,GHz flux density for all the sources in the SERVS fields. The solid diagonal line represents the $R_{3.6}$ detection limit. It is noteworthy that sources not detected in SERVS or with unreliable identifications (triangles) span the entire range in radio flux probed by our samples.

In Table \ref{tab:ratios} we list the number of IFRSs found in each field (CDFS, ELAIS-S1, ELAIS-N1, LH) for different ranges of $R_{3.6}$ ($<$ 100, 100--200, 200--500, and $>$ 500). In case of IR undetected sources (or unreliable identifications) we assign the source to the $R_{3.6}$ range constrained by the estimated lower limit value. Sources with $R_{3.6}$ $<$ 200 cannot be found in the LH field, as we imposed a minimum threshold of 200 for our IFRS search (see Sect.\ \ref{sec:New}). Considering that several sources have $R_{3.6}$ lower limits, at least 60\% of the 55 sources with \mbox{$R_{3.6}$ $>$ 200} satisfy the \citeauthor{Zin2011} $R_{3.6}$ $>$ 500 criterion, and most of them have $S_{1.4\,GHz}$ $>$ 1\,mJy. This is consistent with the fact that this criterion tends to exclude faint radio sources, as discussed in Sect.\ \ref{sec:Intro}. Only three sources in the ELAIS-S1 field show very low radio-to-IR ratios ($R_{3.6}$ $<$ 100) based on our new SERVS analysis, and can be explained in terms of normal galaxy populations. All these sources have \mbox{$S_{1.4\,GHz}$ $<$ 0.2\,mJy}. The rest of the sub-mJy sources (\mbox{$S_{1.4\,GHz}\ \sim$ 0.3--1\,mJy}) typically have $R_{3.6}$ $\sim$ 100--500. They do not satisfy the stringent \citeauthor{Zin2011} \mbox{$R_{3.6}$ $>$ 500} criterion, but display nonetheless extreme IR properties.

\subsection{Average IR flux densities of undetected sources}
\label{subsec:Stack}

We performed a median stacking analysis of the sources with no or unreliable counterparts at the SERVS flux limits, focusing our analysis on IFRSs with \mbox{$R_{3.6}$ $>$ 500} (22 at 3.6\,$\mu$m and 20 at 4.5\,$\mu$m), to explore down to a fainter regime the IR properties of this extreme population. After stacking the SERVS image cutouts, 3.6 and 4.5\,$\mu$m median flux densities were measured through aperture photometry, and flux errors were estimated as explained in Sect.~\ref{sec:Analysis}, except that median stacking removes the need to reject $>$ 3$\sigma$ outliers (see Sect.\ \ref{subsec:Cutouts}).

\begin{table}[!t]
\caption{\label{tab:ratios} Statistics of IFRSs in the SERVS fields as a function of radio-to-infrared flux ratio ($R_{3.6}$). Columns 2, 3, 4, and 5 report the number of sources with $R_{3.6}$ in the ranges $<$ 100, 100--200, 200--500, and $>$ 500, respectively.}
\centering
\renewcommand{\arraystretch}{1.2}
\begin{tabular}{c c c  c c }
\hline\hline
 Field   & \multicolumn{4}{c}{radio--to--infrared flux ratio $R_{3.6}$} \\
         &   $<$ 100   &  100--200  &  200--500  &  $>$ 500  \\
  (1)    &    (2)    &    (3)     &    (4)     &   (5)   \\
\hline
CDFS     &     0     &     3      &      9     &     9   \\
ELAIS-S1 &     3     &     3      &      6     &     5   \\
ELAIS-N1 &     0     &     0      &      0     &     5   \\
LH       &     -     &     -      &      8     &    13   \\
\hline
Total    &     3     &     6      &     23     &    32   \\                                                                                                                                       
\hline
         &           &            &            &         \\
\end{tabular}
\end{table}

No secure detection was obtained in the stacked image. At 3.6\,$\mu$m we find a median flux density upper limit of \mbox{$\widetilde{S}_{3.6\,\mu m}$ $<$ 0.46\,$\mu$Jy}, and for the first time we provide a median upper limit at SERVS 4.5\,$\mu$m, \mbox{$\widetilde{S}_{4.5\,\mu m}$ $<$ 0.60\,$\mu$Jy} (both 3$\sigma$ values). The total lack of detection in the stacked images highlights how the counterparts distribution is probably dominated by sources well below the SERVS detection limit, with actual $R_{3.6}$ values significantly larger than 500.

\citet{Nor2011} also attempted a stacking experiment based on preliminary SERVS images of ATLAS fields, obtaining a median flux density of $\widetilde{S}_{3.6\,\mu m}$ $<$ 0.42\,$\mu$Jy. The two 3.6\,$\mu$m upper limits are remarkably similar despite some of the sources stacked by \citet{Nor2011} are detected at few $\mu$Jy level by us. Our upper limit is slightly larger due to the fact that a smaller number of sources was stacked. The most notable differences between the sample of sources we stacked and the one stacked by \citet{Nor2011}, is that we expanded the sample to two new fields but removed from the stacking all the sources with \mbox{$R_{3.6}$ $<$ 500}. Most of these sources are very faint radio sources (\mbox{$S_{1.4\,GHz}$ $\lesssim$ 0.5\,mJy}), and some have not been detected in ATLAS DR3 \citep{Fra2015}. These sources (CS0275, CS0696, CS0706, CS0714, ES0135, ES1118, and ES1193) may be the result of unusual noise peaks or imaging artefacts.

\begin{table*}[!t]
\caption{\label{tab:ToL} Main modelling parameters}
\centering
\renewcommand{\arraystretch}{1.3}
\begin{adjustbox}{max width=\textwidth}
\begin{tabular}{c c c c c c c c c}
\hline\hline
\multicolumn{8}{c}{\bf{3.6\,\boldmath$\mu$m}} \\
\hline
   \multirow{2}{*}{Class}   &     \multicolumn{2}{c}{\multirow{2}{*}{$L_{z=0}$\,[W/Hz]}}     &                                             \multicolumn{3}{c}{\multirow{2}{*}{Evolution}}                                                                               &    \multirow{2}{*}{$z$ range}     &         \multirow{2}{*}{Bands}          & \multirow{2}{*}{References} \\
                            &                                                                &                                                   &                      &                                                                                               &                                   &                                         & \\
\hline
 \multirow{2}{*}{Arp~220}   & \multicolumn{2}{c}{\multirow{2}{*}{3.25\,$\times$\,10$^{22}$}} &                                                   & \rdelim\{{3}{20pt}[] &\multirow{2}{*}{10$^{0.4\,\left[40.0\ A^{0.19}\ e^{- A } - 22.388 \right]}$}                  &  \multirow{2}{*}{$z$\,$<$\,1.45}    &    \multirow{2}{*}{$L$-, $K$-, $H$-}    & \\
 \multirow{2}{*}{LP/HP RLG} & \multicolumn{2}{c}{\multirow{2}{*}{2.8\,$\times$\,10$^{23}$}}  &         $S_{3.6\,\mu m} \propto$          &                      &\multirow{2}{*}{4.758\,$\times$\,10$^{0.4\,\left[37.6\ B^{0.16}\ e^{- B } - 23.834 \right]}$} &  \multirow{2}{*}{$z \geq$ 1.45}   &    \multirow{2}{*}{$J$-, $I$-, $R$-}    & $^{(a)}$ \\
                            &             &                                                  &                                                   &                      &                                                                                              &                                   &                                         & \\
                            &             &                                                  &                                                   &                      & Where $A = \frac{1 + z}{14.9}$ and $B = \frac{1 + z}{18.5}$                                  &                                   &                                         & \\[2ex]
\cdashline{1-9}   Mrk~231   &           \multicolumn{2}{c}{1.5\,$\times$\,10$^{24}$}         &                                                   &                      &                                                                                              &                                   &                                         & \\
           I19254           &            \multicolumn{2}{c}{5\,$\times$\,10$^{23}$}          & \multirow{2}{*}{$S_{3.6\,\mu m} \propto$} &                      &                     \multirow{2}{*}{10$^{0.8\,z - 0.2\,z^2 + 0.01\,z^3}$}                    &  \multirow{2}{*}{$z$\,$<$\,5.05}    & \multirow{2}{*}{$L$\,$\rightarrow$\,$R$}& \multirow{2}{*}{$^{(b)}$} \\
  Type\,1 QSO/Type\,2 QSO   &           \multicolumn{2}{c}{2.5\,$\times$\,10$^{23}$}         &                                                   &                      &                                                                                              &                                   &                                         & \\
  Type\,1 AGN/Type\,2 AGN   &           \multicolumn{2}{c}{2.5\,$\times$\,10$^{23}$}         &                                                   &                      &                                                                                              &                                   &                                         & \\
\hline\hline
\multicolumn{8}{c}{\bf{4.5\,\boldmath$\mu$m}} \\
\hline
   \multirow{2}{*}{Class}   &     \multicolumn{2}{c}{\multirow{2}{*}{$L_{z=0}$\,[W/Hz]}}     &                                             \multicolumn{3}{c}{\multirow{2}{*}{Evolution}}                                                                              &    \multirow{2}{*}{$z$ range}     &         \multirow{2}{*}{Bands}          & \multirow{2}{*}{References} \\
                            &                                                                &                                                   &                      &                                                                                              &                                   &                                         & \\
\hline
 \multirow{2}{*}{Arp~220}   & \multicolumn{2}{c}{\multirow{2}{*}{4.7\,$\times$\,10$^{22}$}}  &                                                   & \rdelim\{{3}{20pt}[] &\multirow{2}{*}{10$^{0.4\,\left[40.0\ A^{0.19}\ e^{- A } - 22.334 \right]}$}                  &  \multirow{2}{*}{$z$\,$<$\,2.05}    &    \multirow{2}{*}{$L$-, $K$-, $H$-}    & \\
 \multirow{2}{*}{LP/HP RLG} & \multicolumn{2}{c}{\multirow{2}{*}{1.6\,$\times$\,10$^{23}$}}  &         $S_{4.5\,\mu m} \propto$          &                      &\multirow{2}{*}{5.153\,$\times$\,10$^{0.4\,\left[37.6\ B^{0.16}\ e^{- B } - 23.896 \right]}$} &  \multirow{2}{*}{$z \geq$ 2.05}   &    \multirow{2}{*}{$J$-, $I$-, $R$-}    & $^{(a)}$ \\
                            &             &                                                  &                                                   &                      &                                                                                              &                                   &                                         & \\
                            &             &                                                  &                                                   &                      & Where $A = \frac{1 + z}{14.9}$ and $B = \frac{1 + z}{18.5}$                                  &                                   &                                         & \\[2ex]
\cdashline{1-9}   Mrk~231   &          \multicolumn{2}{c}{2.1\,$\times$\,10$^{24}$}          &                                                   &                      &                                                                                              &                                   &                                         & \\
           I19254           &          \multicolumn{2}{c}{8.4\,$\times$\,10$^{23}$}          &                                                   &                      &                                                                                              &                                   &                                         & \\
  Type\,1 QSO/Type\,2 QSO   &          \multicolumn{2}{c}{3.3\,$\times$\,10$^{23}$}          &         $S_{4.5\,\mu m} \propto$          &                      &                              10$^{0.8\,z - 0.2\,z^2 + 0.01\,z^3}$                            &          $z$\,$<$\,5.05             &          $L$\,$\rightarrow$\,$R$        & $^{(b)}$ \\
        Type\,1 AGN         &          \multicolumn{2}{c}{3.6\,$\times$\,10$^{23}$}          &                                                   &                      &                                                                                              &                                   &                                         & \\
        Type\,2 AGN         &          \multicolumn{2}{c}{2.5\,$\times$\,10$^{23}$}          &                                                   &                      &                                                                                              &                                   &                                         & \\
\hline\hline                                                                                                                                                                                                                  
\multicolumn{8}{c}{\bf{1.4\,GHz}} \\
\hline
 \multirow{2}{*}{Class}       &                &     \multirow{2}{*}{$L_{z=0}$\,[W/Hz]}      &                                             \multicolumn{3}{c}{\multirow{2}{*}{Evolution}}                                                             & \multirow{2}{*}{$z_{MAX}$} & \multirow{2}{*}{$\alpha$} & \multirow{2}{*}{References} \\
                              &                &                                             &                                                   &                      &                                                                             &                                   &                           & \\
\hline                                                                                                                               
        Arp~220               &                &            2\,$\times$\,10$^{23}$           &           $S_{1.4\,GHz} \propto$           &                      &                              $(1 + z)^{3.3}$                                &                 2                 &           -1.0            & $^{(c)}$\ $^{(d)}$ \\
                              &                &                                             &                                                   &                      &                                                                             &                                   &                           & \\
\multirow{2}{*}{Ell RLG}      &       LP       &                 10$^{24}$                   &           $S_{1.4\,GHz} \propto$           &                      &                              $(1 + z)^{2.0}$                                &                 2                 &           -1.0            & $^{(e)}$ \\
                              &       HP       &                 10$^{26}$                   &           $S_{1.4\,GHz} \propto$           &                      &                       10$^{(1.26\,z -0.26\,z^2)}$                           &                                   &           -1.0            & $^{(f)}$ \\
                              &                &                                             &                                                   &                      &                                                                             &                                   &                           & \\
         Mrk~231              &                &                 10$^{24}$                   &                                                   &                      &                                                                             &                                   &                           & \\
      Type\,1 AGN             &                &                 10$^{25}$                   &  \multirow{2}{*}{$S_{1.4\,GHz} \propto$}   &                      &                       10$^{(1.18\,z -0.28\,z^2)}$                           &                                   &   \multirow{2}{*}{0.0}    & \multirow{2}{*}{$^{(f)}$} \\
 \multirow{2}{*}{Type\,1 QSO} &       LP       &                 10$^{25}$                   &                                                   &                      &                                                                             &                                   &                           & \\
                              &       HP       &                 10$^{26}$                   &                                                   &                      &                                                                             &                                   &                           & \\
                              &                &                                             &                                                   &                      &                                                                             &                                   &                           & \\
          I19254              &                &                 10$^{24}$                   &                                                   &                      &                                                                             &                                   &                           & \\
      Type\,2 AGN             &                &                 10$^{25}$                   &  \multirow{2}{*}{$S_{1.4\,GHz} \propto$}   &                      &                       10$^{(1.26\,z -0.26\,z^2)}$                           &                                   &   \multirow{2}{*}{-1.0}   & \multirow{2}{*}{$^{(f)}$} \\
 \multirow{2}{*}{Type\,2 QSO} &       LP       &                 10$^{25}$                   &                                                   &                      &                                                                             &                                   &                           & \\
                              &       HP       &                 10$^{26}$                   &                                                   &                      &                                                                             &                                   &                           & \\
\hline
\end{tabular}
\end{adjustbox}
\begin{minipage}{1.0\textwidth}
\small{From \citep{Pol2007}, we used a 5\,Gy old Ell.\ template for the HP/LP RLG track, the QSO1 and QSO2 templates for QSO tracks, and the Seyfert\,1 and 2 templates for AGN tracks; for Arp~220, Mrk~231, and I19254 objects we used the relative templates. $^{(a)}$ \citet{Ste2013}; $^{(b)}$ \citet{Ass2011}; $^{(c)}$ \citet{Hop1998}; \mbox{$^{(d)}$ \citet{Hop2004}}; $^{(e)}$ \citet{Dun1990}.}
\end{minipage}
\end{table*}


\section{Models of comparison}
\label{sec:Models}

Our sample spans a much larger range in both $R_{3.6}$ ratio and radio flux density than the `bright' \citet{Col2014} and \citet{Her2014} samples. Lower flux densities and lower $R_{3.6}$ values can be explained as the result of same IR properties but lower intrinsic radio luminosities, associated with a population of less radio-loud QSOs. Alternatively it can be the result of a more diverse population and/or redshift distribution. Indeed several of our sources lie in the radio sub-mJy regime, where radio sources consist of both AGNs and SFGs (see e.g.\ \citealt{Pra2001, Mig2008, Sey2008, Smo2015}).

To disentangle between these different scenarios, we need to compare the radio/IR properties of our IFRSs with those of known classes of objects, including the effects of evolution and dust extinction.

To build our reference models we used the spectral energy distribution (SED) templates from the SWIRE Template Library \citep{Pol2007}. In particular we used the templates of Arp~220 (as representative of star-burst galaxies), of Mrk~231 (as representative of composite Seyfert\,1/Starburst objects), of IRAS\,19254-7245 (hereafter I19254, as representative of a composite Seyfert\,2/Starburst objects), of a 5 Gyr old Elliptical (for the hosts of elliptical radio-loud galaxies, RLG), Seyfert\,1 and 2 average templates for Type\,1 and 2 AGNs, respectively, and QSO1 and QSO2 templates for average Type\,1 and Type\,2 QSOs.

We included the effect of $K$-correction and intrinsic evolution. For both IR and radio bands, and for all classes of objects, we assumed pure luminosity evolution (PLE), and in particular we used models accounting for a luminosity damping at high redshift ($z$ $\gtrsim$ 2). We $K$-corrected 1.4\,GHz flux densities by assuming a power law spectrum ($S \propto \nu^{\alpha}$), assuming indicative reference values for the spectral index. We assumed $\alpha$ = $-$1.0 for typically steep-spectrum sources (Arp~220, RLGs, I19254, and Type\,2 AGN/QSOs), and $\alpha$ = 0 for typically flat-spectrum sources (Mrk~231 and Type\,1 AGN/QSOs). The 3.6 and 4.5\,$\mu$m flux densities were $K$-corrected using the Hyperz software \citep{Bol2000}.

The intrinsic IR luminosity evolution was modelled following \citet{Ste2013}, who derived PLE models for normal galaxy populations in both rest-frame $H$- and $J$-bands, and following \citet{Ass2011}, who derived PLE models for AGNs in the rest-frame $J$-band. In particular we used the former models for Arp~220 and RLGs, and the latter for Mrk~231, I19254, AGNs and QSOs. The intrinsic IR luminosity evolution for each class of objects was modelled following the evolution of the characteristic luminosity ($L_*$), i.e.\ the luminosity which marks the change from power law to exponential regime in the Schechter luminosity function.

Other models are available in the literature for a number of rest-frame IR bands (see e.g.\ \citealt{Poz2003, Bab2006, Sar2006, Dai2009}). For our toy models, however, subtle differences are not relevant, as we are interested only in obtaining overall reference evolutionary tracks. Our final choice was mainly dictated by: a) the wider redshift range probed by the selected models ($z$ $<$ 3.5 for \citealt{Ste2013}, \mbox{$z$ $<$ 5} for \citealt{Ass2011}), which reduces the uncertainties introduced when such models are extrapolated to higher redshifts; b) the fact that analytical forms were used to describe the evolution that take into account in a single law both positive luminosity evolution at low redshifts and luminosity damping at high redshifts (\mbox{$z$ $\gtrsim$} 2).

For Type\,1 and 2 AGN/QSOs and for RLGs we assumed as reference IR luminosity (\,$L_0 \equiv L_*(z = 0$)\,) the characteristic luminosity expected for these classes of objects at redshift $z$ = 0, following \citet{Ass2011} and \citet{Ste2013}, respectively. For Mrk~231, Arp~220 and I10254 we assumed their own luminosity. In particular we fixed the 3.6\,$\mu$m luminosity and scaled it to 4.5\,$\mu$m following the templates.

We modelled the radio luminosity evolution of high power ($L$ $\gtrsim$ 10$^{25}$\,W\,Hz$^{-1}$) AGNs and composite AGN/starburst galaxies following \citet{Dun1990}. In particular we applied the PLE model derived for steep-spectrum radio sources to high-power RLGs, Type\,2 AGN/QSOs and I19254, and the flat-spectrum model to Type\,1 AGN/QSOs and Mrk~231. Two reference radio powers (\,$L_0 \equiv L(z = 0$)\,) were assumed for RLGs, as well as for Type\,1 and 2 QSOs. For RLGs, the low-power luminosity was assumed 10$^{24}$\,W\,Hz$^{-1}$, while the high-power luminosity is 10$^{26}$\,W\,Hz$^{-1}$. For Type\,1 and 2 QSOs the low-power luminosity is higher than for RLGs (10$^{25}$\,W\,Hz$^{-1}$), while the high-power luminosity is the same (10$^{26}$\,W\,Hz$^{-1}$). For Mrk~231 and IC19254 we assumed 10$^{24}$\,W\,Hz$^{-1}$, which is approximately equal to their actual radio powers.

The radio luminosity of low power ($L$ $\lesssim$ 10$^{24}$\,W\,Hz$^{-1}$) RL AGNs associated to elliptical galaxies is known to evolve less strongly, and is typically modelled with a law of the form $L(z) = L_0 (1 + z)^{\beta}$ up to a given maximum redshift $z_{max}$, and $L(z)= L(z_{max})$ at higher redshifts. Following \citet{Hop2004} we assumed $\beta$ = 2, while $z_{max}$ was set to 2 (this is likely a generous assumption as there are growing indications that the number density of low-power radio loud AGN peaks at redshift \mbox{$z$ $\lesssim$ 1}; see e.g.\ \citealt{Pad2015}).

\begin{figure*}[!t]
\begin{minipage}{0.6\textwidth}
\subfloat[]{\includegraphics[width=300pt]{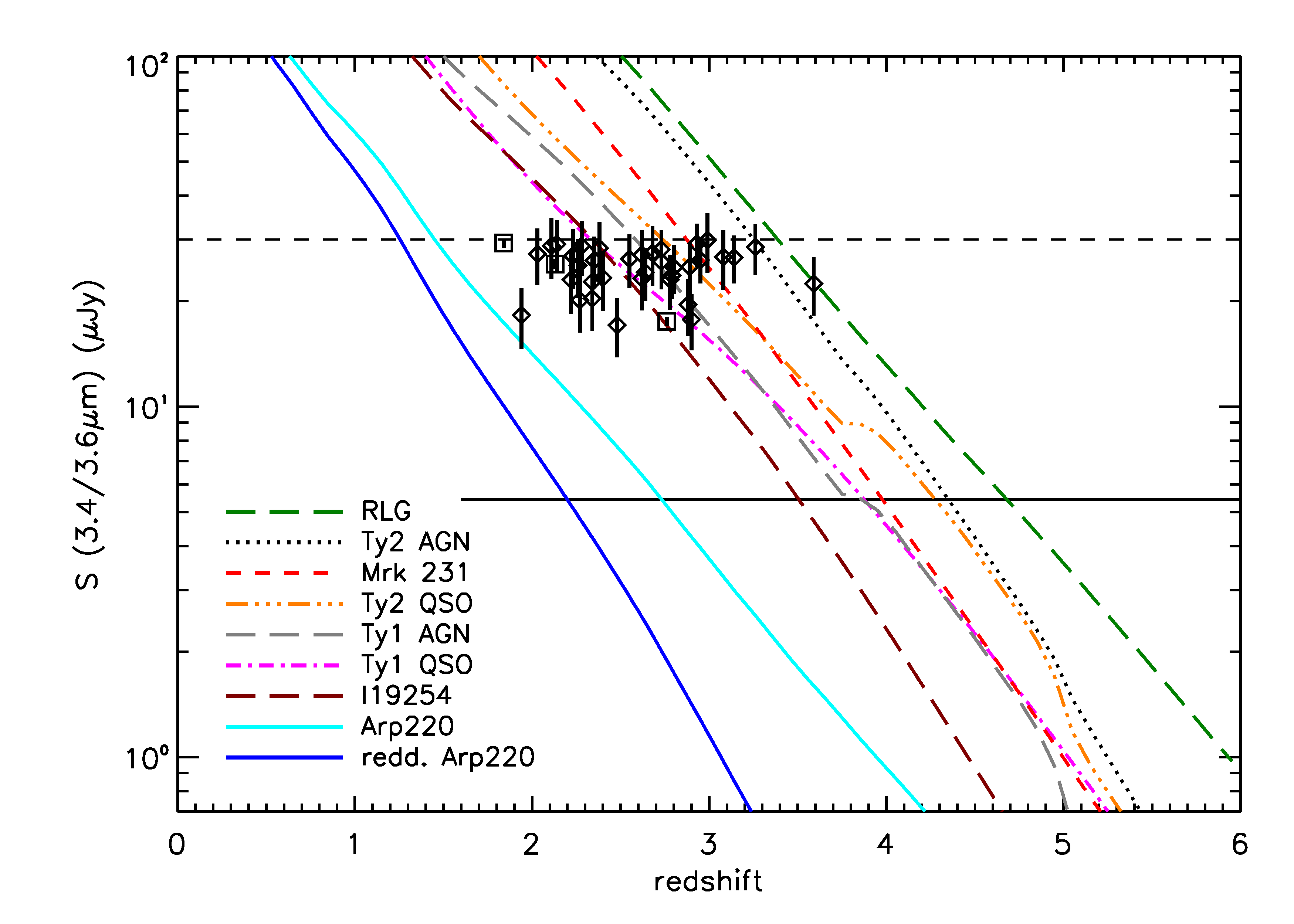}} \\
\subfloat[]{\includegraphics[width=300pt]{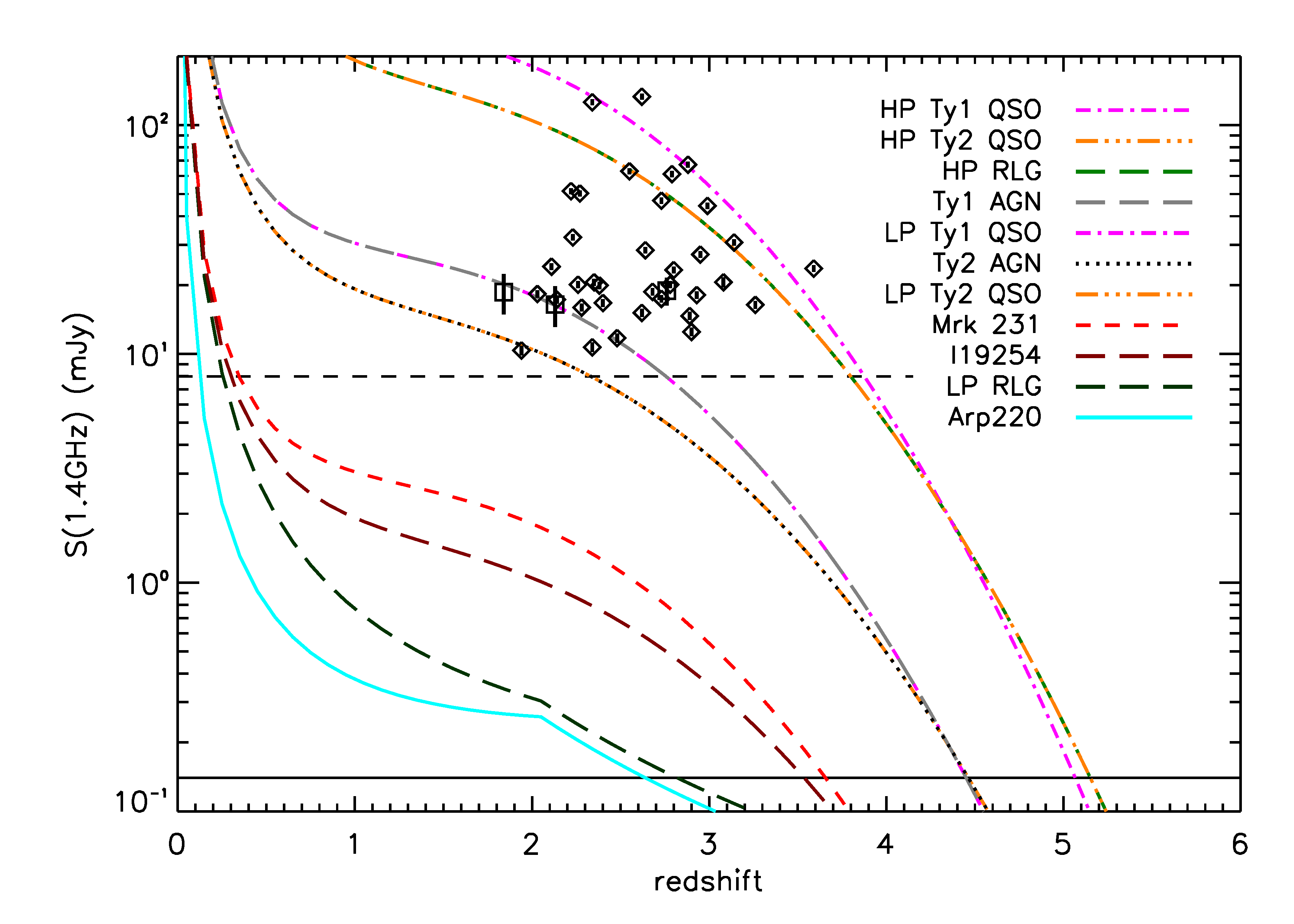}}
\end{minipage} \hspace{0.7cm}
\begin{minipage}{0.33\textwidth}
\caption{\label{fig:Comparison_Models} Expected flux densities versus redshift for our models: Type\,2 and Type\,1 AGNs (dotted black and long-dashed grey tracks, respectively); Type\,2 and Type\,1 QSOs (triple dot-dashed orange and dot-dashed magenta tracks, respectively); radio galaxies (long-dashed green --for high-power-- and dark green --for low-power-- tracks); Arp~220, without and with reddening (dotted cyan and solid blue tracks, respectively), Mrk~231 (dashed red track), and I19254 (long-dashed brown track). Superimposed are the \citet{Col2014} and \citet{Her2014} IFRSs with measured redshift (empty diamonds and empty squares, respectively). {\it Panel (a) -- Top}: IR fluxes at 3.4\,(WISE)/3.6\,(SWIRE) $\mu$m versus redshift; the dashed horizontal line represents the 30\,$\mu$Jy flux density threshold used by \citet{Col2014} and \citet{Her2014} to select their IFRS samples, while the solid horizontal line represents the largest flux density measured in our sample (LH2633, $S_{3.6\,\mu m}$ $\sim$ 5.43\,$\mu$Jy). {\it Panel (b) -- Bottom}: 1.4\,GHz radio flux density versus redshift; high-power Type\,2 QSO and RLG trends are superimposed due to the identical modelling we applied, as well as for Type\,1 AGN and low-power type\,1 QSO, and for Type\,2 AGN and low-power Type\,2 QSO (see Table \ref{tab:ToL}). The dashed horizontal line indicates the lowest 1.4\,GHz flux density measured in \citealt{Col2014} and \citealt{Her2014} samples (7.98\,mJy, see Sect.\ \ref{sec:Intro}), while the solid horizontal line indicates the lowest 1.4\,GHz flux density measured in our sample (ES0463, $S_{1.4\,GHz}$ $\sim$ 0.14\,mJy, see Table \ref{tab:IFRS}).}
\end{minipage}
\end{figure*}

The same evolutionary form is also used to model the radio luminosity of starburst galaxies (Arp~220-like objects). Typically $\beta$ has values between 2.5 and 3.33 (see e.g.\ \citealt{Sau1990, Mac2000, Sad2002, Sey2004, Mao2012}) and $z_{max}$ is typically assumed to be in the range 1.5--2. We assumed $\beta$ = 3.3 and $z_{max}$ = 2 (\citealt{Hop1998, Hop2004}). This is a rough approximation, as it is known that the radio luminosity at high redshift starts to decrease, but this approximation covers well enough the redshift range within which such a source would be still detectable by our radio surveys. The $L_0$ parameter was set equal to the actual 1.4\,GHz radio luminosity of Arp~220 (2 $\times$ 10$^{23}$\,W\,Hz$^{-1}$).

Table \ref{tab:ToL} summarises the reference powers and the evolutionary models applied to our templates, and the redshift range within which these models have been derived.

\begin{figure*}[t!]
\centering
\includegraphics[width=180mm]{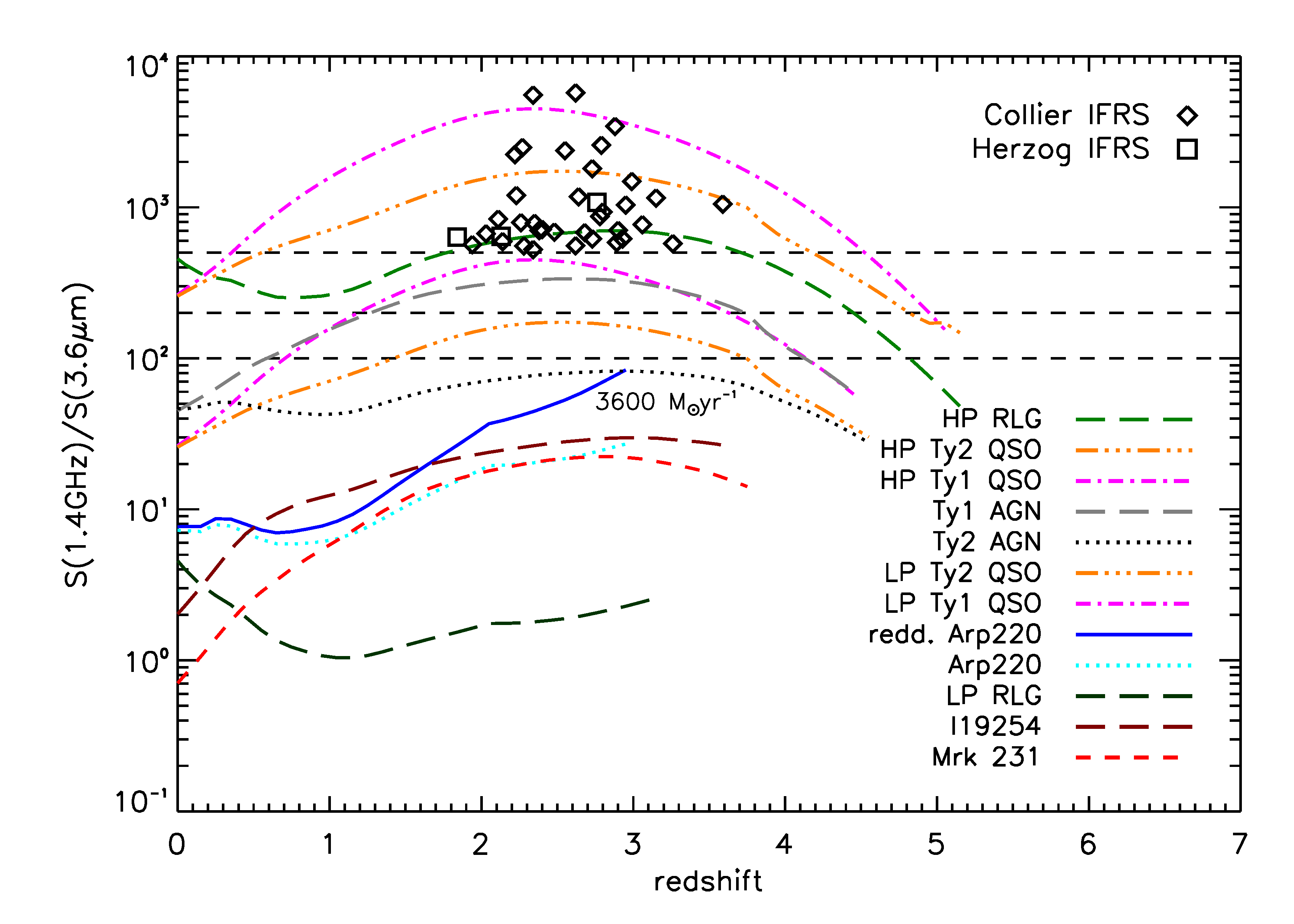}
\caption{\label{fig:RATIO2.png} $R_{3.6}$ ratio evolutionary tracks; in this plot each track is truncated at the redshift at which the radio flux density drops below the detection limit \mbox{$S_{1.4\,GHz} \sim 100\,\mu$Jy}. The horizontal dashed lines indicate $R_{3.6}$ ratios of 100, 200 and 500. The errors associated to the measured $R_{3.6}$ ratios of the \citet{Col2014} and \citet{Her2014} samples are of the same magnitude of the symbols reported in the plot, and not shown. At the end of the Arp~220 reddened track we indicate the star formation rate expected for this source if it were at redshift $\sim$ 3, under the hypothesis that star-forming activity entirely accounts for the radio emission in this object.
}
\end{figure*}

In panels (a), (b) of Figure \ref{fig:Comparison_Models} we show the IR and 1.4\,GHz evolutionary tracks expected for our models, while Figure \ref{fig:RATIO2.png} shows the expected $R_{3.6}$ versus redshift relation. In this case each track is truncated at the redshift at which the radio flux density drops below the typical detection limit of the radio surveys under consideration (\mbox{$S_{1.4\,GHz}$ $\sim$ 0.1\,mJy}, see \mbox{Table \ref{tab:Fields}}). For AGN and QSO templates the expected radio-to-IR $R_{3.6}$ ratio tends to increase with redshift up to $z$ $\sim$ 2--3 due to the stronger positive radio evolution over the IR one. Then it starts to decrease due to the steeper decline in the radio evolution. QSOs, AGNs, and high-power RLGs are the only classes of objects able to reach $R_{3.6}$ $\gtrsim$ 100, further supporting previous evidences that IFRSs host an AGN. Arp~220 (dotted cyan line) keeps increasing up to the highest redshifts at which this source would be still detectable in our surveys, but never reaches the lowest \mbox{$R_{3.6}$ $\gtrsim$ 50} values measured in our faint IFRS sample. The same is true for Mrk~231, low-power RLGs, and I19254.

As pointed out by \citet{Nor2011}, radio-to-IR ratios can be increased by introducing some level of dust extinction. We explored this effect for Arp~220 by modelling the extinction as a power law of the form $A_{\lambda}$ $\propto$ $\lambda^{\alpha}$ \citep{Whi1988}, with $\alpha$ = $-$2.20 (see, e.g., \citealt{Ste2009, Scho2010, Fri2011}). In Figure \ref{fig:Comparison_Models} (panel (a)) and Figure \ref{fig:RATIO2.png} the reddened track of Arp~220 is indicated by the solid blue line. At the end of this track is reported the star formation rate expected for Arp~220 at that redshift (3600\,M$_{\odot}$\,yr$^{-1}$), under the hypothesis that star formation activity entirely accounts for the radio emission in this object. This value has been obtained following \citet{Con1992}. Even in this case starburst Arp~220-like galaxies hardly reach \mbox{$R_{3.6}$ $\sim$ 100}, but can account of the few objects in our IFRS sample with \mbox{$R_{3.6}$ $\lesssim$ 100}.

We notice that the evolutionary tracks shown in panels (a) and (b) of Figure \ref{fig:Comparison_Models} are sensitive to the assumed reference luminosity and therefore should be used with caution. Figure \ref{fig:RATIO2.png}, on the other hand, is more robust as it only depends on the ratio between the radio and IR flux densities. As a sanity check, in both panels of Figure \ref{fig:Comparison_Models} and in Figure \ref{fig:RATIO2.png} we also show the flux densities and/or the $R_{3.6}$ values of all \citet{Col2014} and \citet{Her2014} {\it bright} IFRSs for which a redshift was measured (empty diamonds and empty squares, respectively). All such IFRSs are classified as broad-line Type\,1 QSOs. The only tracks that can reproduce their IR and radio properties (see panels (a) and (b) of Figure \ref{fig:Comparison_Models}), as well as the \mbox{$R_{3.6}$ $\gtrsim$ 500} selection criterion imposed for these IFRSs (see Figure \ref{fig:RATIO2.png}), are the ones of QSOs and HP RLGs. In general, the spectral indices of the \citet{Col2014} IFRSs with measured redshift are flat (see Sect.\ \ref{sec:Intro}), pointing towards a core-dominated Type\,1 QSO population, in excellent agreement with their spectroscopic classification. The two IFRSs from \cite{Her2014} with redshifts, for which a spectral index was measured, show steep spectra ($-$0.84 for CS0212, $-$0.75 for CS0265), possibly indicating that these sources are RL QSO, dominated by optically thin synchrotron emission from the radio jets.

As shown by the horizontal solid lines in Figure \ref{fig:Comparison_Models}, our sample probe much lower radio and IR flux density ranges than the \citet{Col2014} and \citet{Her2014} samples, possibly associated to different source types and/or redshift distribution. This will be investigated in Sect. \ref{sec:Radio/IR_prop}.

\begin{figure*}[!t]
\centering
\includegraphics[width=180mm]{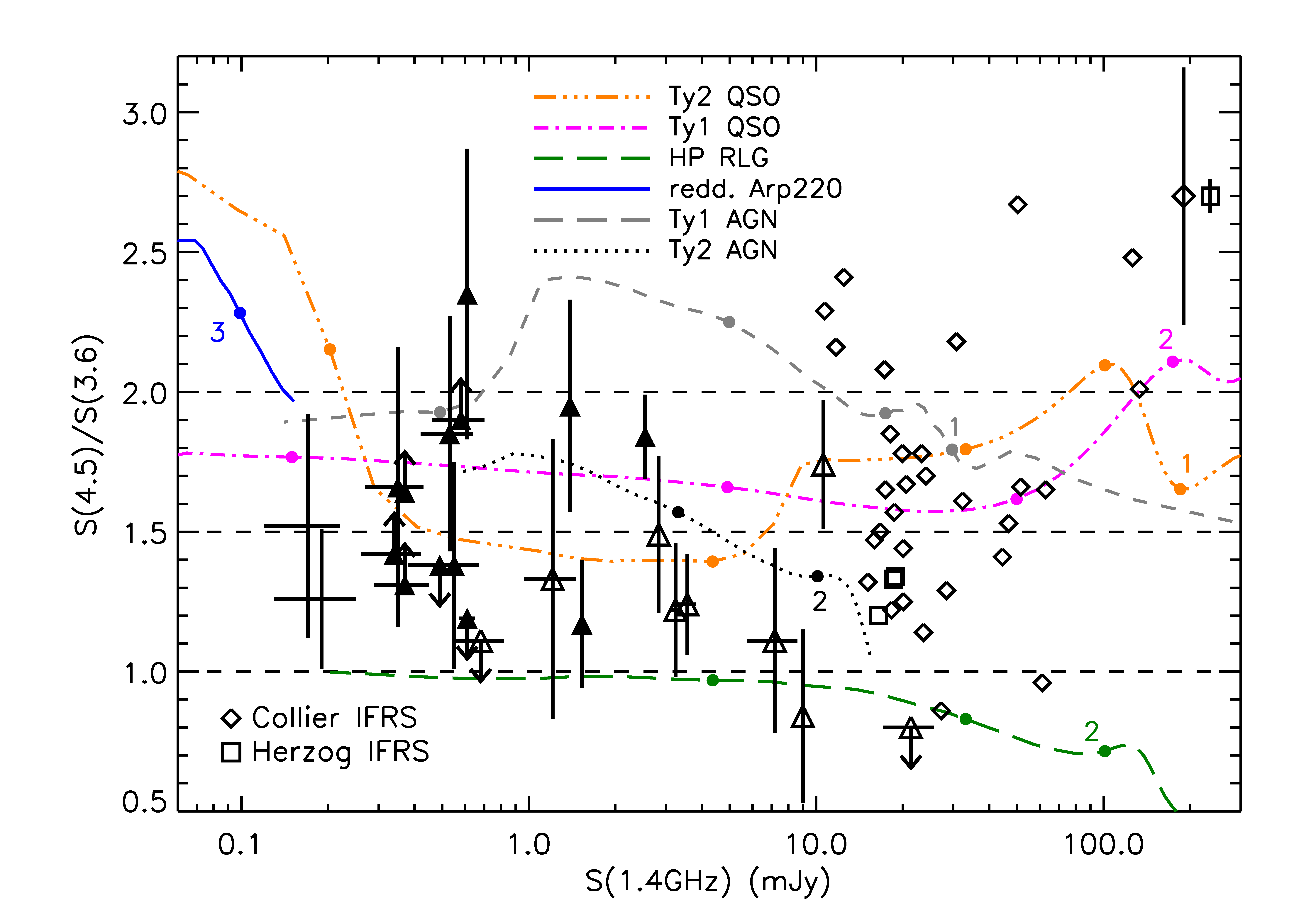}
\caption{\label{fig:Ratio} $S_{4.5\,\mu m}/S_{3.6\,\mu m}$ ratios versus 1.4\,GHz flux density for IFRS sources. Sources detected only at one IR band (either 3.6 or 4.5\,$\mu$m), are indicated by the corresponding upper/lower limits. The dashed horizontal lines refer to $S_{4.5\,\mu m}/S_{3.6\,\mu m}$ ratios equal to 1.0, 1.5 and 2.0 respectively. The superimposed tracks refer to HP Type\,2 and 1 QSOs (triple dot-dashed orange and dot-dashed magenta respectively), HP RLGs (long-dashed green), Type\,1 and 2 AGNs (dashed and dotted grey, respectively) and reddened Arp~220 (solid blue). The redshift increases along the tracks toward the left hand side of the plot; unitary increments in redshift are marked by dots, starting from the value stated in figure. LP QSOs would show a similar track to HP QSOs, but slightly shifted to the left, and are not shown. All the tracks have been drawn in the redshift range within which $R_{3.6}$ $>$ 50, which is close to the smaller value we measured for our IFRSs (see Table \ref{tab:IFRS}). Empty diamonds and empty squares represent the positions of the IFRSs with redshift from \citet{Col2014} and \citet{Her2014}, respectively. In the top-right corner of the plot are shown the median errors for these two samples. Upward empty and filled triangles indicate IFRSs from our sample with $R_{3.6}$ $>$ 500 and between 200 and 500, respectively. The two radio-faintest sources are ES0593 and ES0436, both with $R_{3.6}$ values below 100. None of the reliably identified IFRSs shown in this plot has $R_{3.6}$ between 100 and 200.}
\end{figure*}

\section{Radio/IR properties of SERVS deep field samples}
\label{sec:Radio/IR_prop}

We used the evolutionary tracks described in Sect.\ \ref{sec:Models} to assess the nature of our $\gtrsim$ 10$\times$ fainter IFRSs, spanning a larger range of $R_{3.6}$ values (i.e.\ $R_{3.6}$ $\gtrsim$ 50--100). In absence of redshift information, we explored the radio and IR properties of our sample in the parameter space defined by the $S_{4.5\,\mu m}/S_{3.6\,\mu m}$ flux density ratio against the 1.4\,GHz radio flux. This choice has the advantage of being independent of the assumed reference IR luminosity, while different radio luminosities just produce an horizontal shift of the evolutionary tracks.

The results are shown in Figure \ref{fig:Ratio}, in which we plot only those IFRSs detected at either 3.6 or 4.5\,$\mu$m, or both. Our IFRSs typically have $S_{4.5\,\mu m}/S_{3.6\,\mu m}$ flux ratios in the range 1--2. Due to the very large error bars ($\widetilde{\epsilon}$ = 0.35 for our sample) it is difficult to say if any source has the most extreme values $S_{4.5\,\mu m}/S_{3.6\,\mu m} \geq$ 2. Different symbols correspond to different $R_{3.6}$ ranges: $R_{3.6}$ $>$ 500 (empty triangles), 200 $<$ $R_{3.6}$ $<$ 500 (filled triangles), $R_{3.6}$ $<$ 100 (error bars only). None of the reliably identified IFRS shown in Figure \ref{fig:Ratio} has 100 $<$ $R_{3.6}$ $<$ 200.

Superimposed are the evolutionary tracks of the classes discussed in Sect.\ \ref{sec:Models} (following the same colour and line style convention as in Figure \ref{fig:Comparison_Models}), that can produce $R_{3.6}$ $>$ 50 (namely Type\,1 and 2 QSO, HP RLGs, Type\,1 and 2 AGNs and reddened Arp~220). The redshift increases along the tracks toward the left hand side of the plot, and each unitary increment of redshift is marked with a dot of the same colour of the track. The first of these dots reports also the first unit redshift of that track. Overall, the $S_{4.5\,\mu m}/S_{3.6\,\mu m}$ flux ratios spanned by our models cover the entire range spanned by our IFRS sample, indicating that our models can account for the whole faint IFRS population.

In Figure \ref{fig:Ratio} we also show the brighter IFRSs from \citet{Col2014} and \citet{Her2014}. The median errors on their IR ratios are shown in the upper right corner of the figure (\mbox{$\widetilde{\epsilon}$ = 0.46} and = 0.06 for \citealt{Col2014} and \citealt{Her2014}, respectively). Considering the large error bars all these IFRSs are again consistent with a QSO classification. Our fainter IFRS sources are mostly consistent with being higher redshift counterparts of \citet{Col2014} and \citet{Her2014} IFRSs. Sources with $R_{3.6}$ $>$ 500 would be QSOs at redshifts (\mbox{$z$ $\sim$ 3--4}), while IFRSs with less extreme $R_{3.6}$ \mbox{(100 $<$ $R_{3.6}$ $<$ 500)} would lie at higher redshift (\mbox{$z$ $>$ 4}). This is consistent with the expected $R_{3.6}$ vs.\ redshift relation of QSOs reported in Figure \ref{fig:RATIO2.png}. Only the IFRSs with lowest $S_{4.5\,\mu m}/S_{3.6\,\mu m}$ flux ratios ($\sim$ 1) are better described by the HP RLG track, and their $R_{3.6}$ values are again consistent with the expected $R_{3.6}$ vs.\ redshift relation.

We notice that Type\,1 and 2 AGN tracks can also account for $R_{3.6}$ $<$ 500 IFRSs. Therefore, the fainter IFRS population could in principle either be associated to very high redshift QSOs or to Type\,1 and 2 AGNs at less extreme redshift (2 $<$ $z$ $<$ 4), or a mixture of both.

Our faint IFRS sources have average radio spectral properties in line with \citet{Her2014} IFRSs. \citet{Fra2015} reported the spectral indices computed between 1.40 and 1.71\,GHz, values that are consistent with the ones computed between 1.4\,GHz (from low-resolution data) and 2.3\,GHz by \citet{Zin2012}, and with the ones reported by \citet{Mid2008a}. Twenty-one of our sources in the CDFS and ELAIS-S1 have measured spectral indices (see Table \ref{tab:IFRS}, Col.\ 16). Thirteen of them have $R_{3.6}$ $>$ 500, and their median spectral index is very steep (\mbox{$\widetilde{\alpha}$ = $-$1.05}), while 8 of them have 200 $<$ $R_{3.6}$ $<$ 500, and their median spectral index is (\mbox{$\widetilde{\alpha}$ = $-$0.72}). A steep-spectrum population is more consistent with a Type\,2 QSO/AGN or RLG classification. 

As a final remark we notice that the faintest IFRSs in our sample ($S_{1.4\,GHz}$ $\lesssim$ 200\,$\mu$Jy), characterized by very low $R_{3.6}$ values ($<$ 100), could also be associated with heavily obscured dust-enschrouded starburst galaxies at medium-high redshift ($z$ $\sim$ 2--3; blue track in Figure \ref{fig:Ratio}), even though in this case we would expect $S_{4.5\,\mu m}/S_{3.6\,\mu m}$ $\gtrsim$ 2.


\section{Summary \& Conclusions}
\label{sec:Conclusion}

In this paper we presented a new study of the radio and IR properties of the IFRSs originally discovered in a number of the SWIRE fields, based on the deeper SERVS images and catalogues now available. This study was complemented by a new IFRS sample extracted in the LH region covered by SERVS.

We repeated the analysis performed by \citet{Nor2011} on the 3.6\,$\mu$m SERVS images of the CDFS and ELAIS-S1 fields, using more recent and deeper 3.6\,$\mu$m SERVS data, and for the first time 4.5\,$\mu$m SERVS images. In addition we extended the analysis to the existing IFRS SWIRE-based sample in the ELAIS-N1 field. For the LH field, we extracted a new sample directly using the SERVS data. This sample consists of 21 new IFRSs.

Most of our sources are characterised by \mbox{0.1\,mJy $<$ $S_{1.4\,GHz}$ $<$ 10\,mJy}. Thanks to the new deeper SERVS images and the use of the likelihood ratio cross-matching, we significantly increased the number of sources detected at 3.6\,$\mu$m (with respect both SWIRE and SERVS pre-release surveys; see Sect.\ \ref{subsec:Stack}). In addition we provided 4.5\,$\mu$m flux density measurements for 25 objects. We identified 21 reliable counterparts at 3.6\,$\mu$m and 20 at 4.5\,$\mu$m, for a total of 25 distinct IFRSs. Sixteen of them have been detected in both bands. Most of the identified IFRS sources have IR fluxes of a few $\mu$Jy, typically corresponding to \mbox{S/N $\sim$ 3--10} in the SERVS images.

Given the different selection criteria used to identify the original IFRSs, the radio-to-IR ratio range spanned by them is rather large. From our new analysis we found that a couple of the original \citet{Nor2006} and \citet{Mid2008a} IFRSs have $R_{3.6}$ $<$ 100, where contamination from intermediate redshift (dust-enshrouded) star-burst galaxies is expected. In other cases, the new, fainter 3.6\,$\mu$m upper limits that we derived confirmed we are dealing with IFRSs characterized up to extremely large $R_{3.6}$ values ($\gg$ 500).
We compared the observational radio/IR properties of our sample, as well as of the brighter IFRS samples extracted by \citet{Col2014} and \citet{Her2014} (\mbox{$S_{1.4\,GHz}$ $\sim$ 8.00 $\rightarrow$ 800\,mJy} for \citealt{Col2014}; \mbox{$S_{1.4\,GHz}$ $\sim$ 7.00 $\rightarrow$ 26.00\,mJy} for \citealt{Her2014,Her2015b}), with those expected for a number of known prototypical classes of objects: Arp~220 (as starburst galaxy), high-power and low-power radio-loud galaxies (RLG), Type\,1 and 2 (i.e.\ obscured) QSOs and AGNs, Mrk~231 (for a prototypical Seyfert\,1/Starburst composite galaxy) and I19254 (for a prototypical Seyfert\,2/Starburst composite galaxy). For each class we built evolutionary models taking into account $K$-correction and evolution, both in IR and radio bands. In the IR domain, we made the simplified assumption that these classes of sources evolve following the characteristic luminosity of the Schechter luminosity function. 

In general we found that the only evolutionary tracks that can produce high $R_{3.6}$ values ($\gtrsim$ 100) are those of AGN-driven sources (AGNs, QSOs, and powerful RLGs). In addition we found that the predicted $R_{3.6}$ values typically show a peak at redshift $z$ 2--3, where $R_{3.6}$ can be larger than 500. Then the expected $R_{3.6}$ decreases to smaller values ($>$ 100--200 at $z$ $\sim$ 5). The radio/IR properties and redshift distribution of the bright, \mbox{$R_{3.6}$ $>$ 500}, IFRSs selected by \citet{Col2014} and \citet{Her2014}, are well reproduced by QSO evolutionary tracks, in excellent agreement with the Type\,1 QSO optical classification of those with spectroscopy available.

In absence of redshift information we analysed the radio/IR properties of the fainter IFRSs in the SERVS deep fields in the parameter space defined by the (luminosity independent) $S_{4.5\,\mu m}/S_{3.6\,\mu m}$ flux density ratio against the 1.4\,GHz radio flux. We found that most of our sources are consistent with being either very high redshift ($z$ $>$ 4) QSOs or Type\,1/2 AGNs at less extreme redshifts (2 $<$ $z$ $<$ 4). Their steep radio spectral indices seem more consistent with a Type\,2 population. Only those IFRSs characterized by low $S_{4.5\,\mu m}/S_{3.6\,\mu m}$ values ($\sim$ 1) are better reproduced by powerful RLG evolutionary tracks (also characterized by steep spectra). Overall the $R_{3.6}$ values of our IFRSs are in good agreement with the redshift distribution predicted by the AGN/QSO and high-power RLG evolutionary tracks, and less extreme ($R_{3.6}$ $\sim$ 200) IFRS sources could be a mixture of higher redshift and/or lower luminosity counterparts of the \citet{Col2014} and \citet{Her2014} bright IFRS samples. Only spectroscopic follow-ups can disentangle between these two alternative scenarios.

Nevertheless, we cannot exclude that the two faintest (\mbox{$S_{1.4\,GHz}$ $\lesssim$ 0.2\,mJy}) sources, both with $R_{3.6}$ $<$ 100, could be associated with heavily obscured dust-enshrouded starburst galaxies, even though in this case larger $S_{4.5\,\mu m}/S_{3.6\,\mu m}$ values than those observed are predicted ($>$ 2). Under this hypothesis, these two sources would lie at redshift $\sim$ 2.4, have a radio power \mbox{$L$ $\sim$ 9 $\times$ 10$^{24}$\,W\,Hz$^{-1}$}, and a SFR $\gtrsim$ 4 $\times$ 10$^3$\,M$_{\odot}$\,yr$^{-1}$. These values are high, but observed in HyLIRG and/or Sub-Millimeter Galaxies (SMG) (see e.g.\ \citealt{Al-Her2013} and \citealt{Bar2014}).

Finding very high-redshift radio-loud AGNs is of great importance to understand the formation and evolution of the first generations of super-massive black holes. Only a few $z$ $>$ 5 radio sources have been found so far (the highest redshift radio galaxy being at $z$ $\sim$ 5.19; \citealt{VanB1999}), and a number of techniques to efficiently pre-select high-redshift candidates are being used (see e.g.\ \citealt{Fal2004}).

As suggested by \citet{Nor2011} and \citet{Her2014}, a $S_{3.6\,\mu m}$ versus $z$ relation seems to exist for \citeauthor{Zin2011} IFRSs, and the radio/IR properties of our fainter sample support that conclusion. Coupled with the radio-to-IR ratio ($R_{3.6}$) criterion, these relations can provide  efficient criteria to pinpoint very high redshift RL QSOs and/or RLGs. Since the radio evolution of such sources seems to be stronger than the IR one, imposing a threshold of $R_{3.6}$ $>$ 500 to select IFRSs, would  result in rejecting the highest-redshift tail of these sources (i.e., IFRSs at redshift \mbox{$z$ $\gtrsim$ 4}). On the other hand, at 1.4\,GHz flux densities below 200\,$\mu$Jy, some contamination from intermediate-redshift \mbox{($z$ $\sim$ 2--3)} dust-enshrouded starburst galaxies is expected. To minimize such contamination only sources with $R_{3.6}$ $>$ 100--150 should be retained as high-redshift IFRS candidates.

\begin{acknowledgements}
The authors thanks the anonymous referee for the valuable comments, that allowed us to improve the discussion of our results.

AM is responsible for the content of this publication.

AM acknowledges funding by a Cotutelle International Macquarie University Research Excellence Scholarship (iMQRES).

IP and RPN acknowledge support from the Ministry of Foreign Affairs and International Cooperation, Directorate General for the Country Promotion (Bilateral Grant Agreement ZA14GR02 - Mapping the Universe on the Pathway to SKA).

RM gratefully acknowledge support from the European Research Council under the European Union's Seventh Framework Programme (FP/2007-2013) /ERC Advanced Grant RADIOLIFE-320745.

This work is based in part on observations made with the Spitzer Space Telescope, which is operated by the Jet Propulsion Laboratory, California Institute of Technology under a contract with NASA.

IRAF is distributed by the National Optical Astronomy Observatory, which is operated by the Association of Universities for Research in Astronomy (AURA) under cooperative agreement with the National Science Foundation.

This research has made use of NASA's Astrophysics Data System.

This research has made use of the Ned Wright's Javascript Cosmology Calculator \citep{Wri2006}.
\end{acknowledgements}

    \bibliographystyle{aa} 
    \bibliography{Biblio} 

\clearpage

\begin{landscape}
\begin{table}[!t]
\caption{\label{tab:IFRS} Radio and infrared properties of the candidate IFRSs in SERVS fields. Column (1): field name; Column (2): source identifier; Columns (3) and (4): position of the radio source; Columns (5) and (6): uncertainties in the radio position; Column (7) and (8): distance from the claimed IR counterpart; Column (9): reliability of the identification (this value refers to the cross-matched 3.6\,$\mu$m counterpart, unless the counterpart has been detected only at 4.5\,$\mu$m); Column (10): 1.4\,GHz flux density; Columns (11) and (12): measured 3.6 and 4.5\,$\mu$m flux densities (3$\sigma$ upper limits are provided in case of no detection); Columns (13) and (14): 1.4\,GHz-to-3.6\,$\mu$m and 1.4\,GHz-to-4.5\,$\mu$m flux ratios, respectively (3$\sigma$ lower limits are provided in case of no detection); Column (15): 4.5\,$\mu$m-to-3.6\,$\mu$m flux ratio (3$\sigma$ upper/lower limits are provided in case of detection at only one band); Column (16): Spectral index (from \citealt{Fra2015}, except for ES0973 from \citealt{Mid2011}); Column (17): Optical magnitude (AB system; band of reference as subscript). We notice that in case of unreliable IR identifications we provide both the measured estimates and 3$\sigma$ limits (following row) for all IR or IR-derived parameters.}
\scriptsize
\centering
\renewcommand{\arraystretch}{1.2}
\addtolength{\tabcolsep}{-2pt}
\begin{adjustbox}{min width=\textwidth}
\begin{tabular}{c c c c c c c c c c c c c c c c c}
\hline\hline
Field & IFRS & \multicolumn{2}{c}{Radio coordinates} & \multicolumn{2}{c}{Combined positional} &      \multicolumn{2}{c}{Distance from claimed IR}       & Reliability &  $S_{1.4\,GHz}$  & $S_{3.6\,\mu m}$ & $S_{4.5\,\mu m}$ &      $S_{1.4\,GHz}/S_{3.6\,\mu m}$     &      $S_{1.4\,GHz}/S_{4.5\,\mu m}$     &$S_{4.5\,\mu m}/S_{3.6\,\mu m}$& Spectral &  Optical  \\
                                                                                                                                                                                                                                             
name  & ID   &      \multicolumn{2}{c}{(J2000)}      &  \multicolumn{2}{c}{accuracy (arcsec)}  & \multicolumn{2}{c}{3.6/4.5\,$\mu$m couterpart (arcsec)} &             &                         &                          &                          &$S_{1.4\,GHz}/(3 \times N_{3.6\,\mu m}$)&$S_{1.4\,GHz}/(3 \times N_{4.5\,\mu m})$&                                               &  index   & magnitude \\
                                                                                                                                                                                                                                                                                                                                                                                                    
      &      &    (hh:mm:ss)    &    (dd:mm:ss)      &   $\sigma_{RA}$   &   $\sigma_{Dec}$    &          $\Delta$RA        &           $\Delta$Dec      &     (\%)    &          (mJy)          &        ($\mu$Jy)         &       ($\mu$Jy)          &                                                       &                                                       &                                               &          &           \\
                                                                                                                                                                                                                                                                                                                                                                                                    
 (1)  & (2)  &       (3)        &      (4)           &         (5)       &         (6)         &             (7)            &             (8)            &     (9)     &          (10)           &           (11)           &           (12)           &                          (13)                         &                         (14)                          &                      (15)                     &   (16)   &    (17)   \\
\hline
 & CS0114  & 03:27:59.89 & $-$27:55:54.7 & 1.02 & 1.02 & 0.12 & 0.89 & 99 & 7.17 $\pm$ 1.43  & 2.13 $\pm$ 0.47  & 2.37 $\pm$ 0.47  & 3366 $\pm$ 1001     & 3025 $\pm$ 851      & 1.11 $\pm$ 0.33 &     $-$1.33    & \\*
 & CS0122  & 03:28:12.99 & $-$27:19:42.6 & 1.42 & 3.59 & 1.90 &      & 86 & 0.46 $\pm$ 0.10  & 1.47 $\pm$ 0.46  & $<$ 1.29           & 313 $\pm$ 119       & $>$ 357               & $<$ 0.88          &                & \\*
 &         &             &               &      &      &      &      &    &                  & $<$1.38            & $<$ 1.29           & $>$ 333               & $>$ 357               & \dots           &                & \\*
 & CS0164  & 03:29:00.20 & $-$27:37:45.7 & 1.02 & 1.26 & 1.21 & 1.06 & 95 & 1.21 $\pm$ 0.25  & 1.89 $\pm$ 0.50  & 2.51 $\pm$ 0.67  & 640 $\pm$ 215       & 482 $\pm$ 163       & 1.33 $\pm$ 0.50 &     $-$0.26    & \\*
 & CS0173  & 03:29:09.66 & $-$27:30:13.7 & 1.42 & 3.77 & 1.71 & 1.66 & 96 & 0.35 $\pm$ 0.08  & 1.66 $\pm$ 0.41  & 2.76 $\pm$ 0.48  & 211 $\pm$ 71        & 127 $\pm$ 36        & 1.66 $\pm$ 0.50 &     $-$0.01    & \\*
 & CS0194  & 03:29:28.59 & $-$28:36:18.8 & 1.02 & 1.02 &      &      &    & 6.09 $\pm$ 1.22  & $<$ 1.77           & $<$ 2.16           & $>$ 3441              & $>$ 2819              & \dots           &     $-$0.92    & \\*
 & CS0215  & 03:29:50.01 & $-$27:31:52.6 & 1.02 & 1.02 &      &      &    & 1.10 $\pm$ 0.22  & $<$ 1.50           & $<$ 1.50           & $>$ 733               & $>$ 733               & \dots           &     $-$0.71    & \\*
 & CS0241  & 03:30:10.21 & $-$28:26:53.0 & 1.02 & 1.54 &      &      &    & 1.28 $\pm$ 0.26  & $<$ 1.41           & $<$ 1.71           & $>$ 908               & $>$ 749               & \dots           &     $-$1.05    & \\*
 & CS0255  & 03:30:24.08 & $-$27:56:58.7 & 1.20 & 3.17 & 0.36 & 0.85 & 99 & 0.55 $\pm$ 0.12  & 2.42 $\pm$ 0.50  & 3.35 $\pm$ 0.56  & 227 $\pm$ 68        & 164 $\pm$ 45        & 1.38 $\pm$ 0.37 &       0.04     & \\*
 & CS0275  & 03:30:43.69 & $-$28:47:55.6 & 1.39 & 1.83 &      &      &    & 0.36 $\pm$ 0.08  & $<$ 1.20           & $<$ 1.68           & $>$ 300               & $>$ 214               & \dots           &                & \\*
 & CS0283  & 03:30:48.68 & $-$27:44:45.3 & 1.17 & 2.04 &  --  &  --  & -- & 0.29 $\pm$ 0.07  &        --        &        --        &          --         &         --          &        --       &       --       & -- \\*
 & CS0415  & 03:32:13.07 & $-$27:43:51.0 & 1.02 & 1.02 &      &      &    & 1.21 $\pm$ 0.25  & $<$ 1.02           & $<$ 1.20           & $>$ 1186              & $>$ 1008              & \dots           & -1.19          & \\*
 & CS0446  & 03:32:31.54 & $-$28:04:33.5 & 2.22 & 3.65 & 2.00 &      & 87 & 0.34 $\pm$ 0.08  & 1.55 $\pm$ 0.50  & $<$ 2.07           & 219 $\pm$ 88        & $>$ 164               & $<$ 1.34          &                & 25.55$_B$ $^{(a)}$ \\*
 &         &             &               &      &      &      &      &    &                  & $<$ 1.50           & $<$ 2.07           & $>$227                & $>$ 164               & \dots           &                & \\*
 & CS0487  & 03:33:01.19 & $-$28:47:20.7 & 1.02 & 1.60 &      &      &    & 1.12 $\pm$ 0.23  & $<$ 1.62           & $<$ 2.01           & $>$ 691               & $>$ 557               & \dots           &     $-$0.62    & \\*
\multirow{2}{*}{\rotatebox{90}{CDFS}} & CS0506  & 03:33:11.48 & $-$28:03:19.0 & 1.57 & 2.37 & 2.44 & 2.57 & 83 & 0.17 $\pm$ 0.06  & 1.80 $\pm$ 0.47  & 2.20 $\pm$ 0.50  & 94 $\pm$ 41         & 77 $\pm$ 32         & 1.22 $\pm$ 0.42 &              & 26.16$_B$ $^{(a)}$ \\*
 &         &             &               &      &      &      &      &    &                  &                  &                  &                     &                     &                 &                & 27.10$_U$ $^{(b)}$ \\*
 &         &             &               &      &      &      &      &    &                  &                  &                  &                     &                     &                 &                & 26.15$_B$ $^{(b)}$ \\* 
 &         &             &               &      &      &      &      &    &                  &                  &                  &                     &                     &                 &                & 26.59$_V$ $^{(b)}$ \\* 
 &         &             &               &      &      &      &      &    &                  &                  &                  &                     &                     &                 &                & 26.16$_R$ $^{(b)}$ \\* 
 &         &             &               &      &      &      &      &    &                  &                  &                  &                     &                     &                 &                & 26.06$_I$ $^{(b)}$ \\* 
 &         &             &               &      &      &      &      &    &                  &                  &                  &                     &                     &                 &                & 24.38$_Z$ $^{(b)}$ \\*
 &         &             &               &      &      &      &      &    &                  & $<$ 1.41           & $<$ 1.50           & $>$ 121               & $>$ 113               & \dots           &                & \\*
 & CS0538  & 03:33:30.20 & $-$28:35:11.1 & 1.45 & 2.58 &      &      &    & 1.40 $\pm$ 0.28  & $<$ 2.16           & $<$ 1.95           & $>$ 648               & $>$ 718               & \dots           &     $-$1.19    & \\*
 & CS0588  & 03:34:04.70 & $-$28:45:01.7 & 1.30 & 3.32 &      &      &    & 0.45 $\pm$ 0.10  & $<$ 1.41           & $<$ 1.50           & $>$ 319               & $>$ 300               & \dots           &                & \\*
 & CS0682  & 03:35:18.48 & $-$27:57:42.2 & 1.20 & 2.88 &      & 0.73 & 96 & 0.34 $\pm$ 0.08  & $<$ 1.59           & 2.26 $\pm$ 0.72  & $>$ 214               & 150 $\pm$ 60        & $>$ 1.42          &                & \\*
 & CS0694  & 03:35:25.08 & $-$27:33:13.2 & 1.02 & 1.40 &      &      &    & 0.60 $\pm$ 0.13  & $<$ 1.98           & $<$ 2.31           & $>$ 303               & $>$ 260               & \dots           &     $-$0.93    & 24.91$_i$ $^{(c)}$ \\* 
 & CS0696  & 03:35:25.25 & $-$28:31:05.2 & 1.02 & 1.78 &      &      &    & 0.31 $\pm$ 0.07  & $<$ 1.65           & $<$ 1.71           & $>$ 188               & $>$ 181               & \dots           &                & \\*
 & CS0703  & 03:35:31.02 & $-$27:27:02.2 & 1.02 & 1.02 & 3.12 & 3.27 & 70 & 26.08 $\pm$ 5.22 & 2.34 $\pm$ 0.61  & 1.97 $\pm$ 0.51  & 11145 $\pm$ 3663    & 13239 $\pm$ 4332    & 0.84 $\pm$ 0.31 &     $-$0.96    & \\*
 &         &             &               &      &      &      &      &    &                  & $<$ 1.83           & $<$ 1.53           & $>$ 14251             & $>$ 17046             & \dots           &                & \\*
 & CS0706  & 03:35:33.22 & $-$28:06:21.8 & 1.07 & 1.51 &      &      &    & 0.26 $\pm$ 0.07  & $<$ 1.62           & $<$ 1.95           & $>$ 160               & $>$ 133               & \dots           &                & \\*
 & CS0714  & 03:35:38.16 & $-$27:44:00.6 & 1.45 & 2.22 &      &      &    & 0.39 $\pm$ 0.09  & $<$ 1.23           & $<$ 1.02           & $>$ 317               & $>$ 382               & \dots           &                & \\*
 &         &             &               &      &      &      &      &    &                  &                  &                  &                     &                     &                 &                & \\*
 & ES0056  & 00:33:46.75 & $-$44:29:02.8 & 1.02 & 1.19 &      & 0.43 & 99 & 0.58 $\pm$ 0.12  & $<$ 1.56           & 2.97 $\pm$ 0.80  & $>$ 372               & 195 $\pm$ 66        & $>$ 1.90          &     $-$2.35    & \\*
 & ES0135  & 00:33:30.12 & $-$44:21:15.4 & 1.60 & 1.56 &      &      &    & 0.18 $\pm$ 0.05  & $<$ 2.28           & out              & $>$ 79                & \dots               & \dots           &                & \\*
\multirow{3}{*}{\rotatebox{90}{ELAIS-S1}} & ES0156  & 00:34:46.40 & $-$44:19:26.9 & 1.02 & 1.41 &      & 0.14 & 99 & 0.37 $\pm$ 0.08  & $<$ 1.53           & 2.00 $\pm$ 0.44  & $>$ 242               & 185 $\pm$ 18        & $>$ 1.31          &     $-$0.88    & \\*
 & ES0318  & 00:37:05.54 & $-$44:07:33.6 & 1.02 & 1.14 &      &      &    & 1.59 $\pm$ 0.32  & $<$ 1.65           & $<$ 1.98           & $>$ 964               & $>$ 803               & \dots           &     $-$1.42    & \\*
 & ES0427  & 00:34:11.59 & $-$43:58:17.0 & 1.02 & 1.02 & 0.17 &      & 99 & 21.36 $\pm$ 4.27 & 1.94 $\pm$ 0.46  & $<$ 1.56           & 11010 $\pm$ 3415    & $>$ 13692             & $<$ 0.80          &     $-$0.95    & \\*
 & ES0433  & 00:34:13.43 & $-$43:58:02.4 & 1.02 & 1.02 &      &      &    & 0.25 $\pm$ 0.06  & $<$ 1.62           & $<$ 1.41           & $>$ 154               & $>$ 177               & \dots           &     $-$1.41    & \\*
 & ES0436  & 00:37:26.34 & $-$43:57:33.0 & 1.09 & 1.54 & 1.20 & 1.07 & 93 & 0.19 $\pm$ 0.06  & 3.27 $\pm$ 0.42  & 4.12 $\pm$ 0.64  & 58 $\pm$ 20         & 46 $\pm$ 16         & 1.26 $\pm$ 0.25 &                & \\
\hline
\end{tabular}
\end{adjustbox}
\end{table}
\end{landscape}

\clearpage

\begin{landscape}
\begin{table}[!t]
\scriptsize
\centering
\renewcommand{\arraystretch}{1.2}
\addtolength{\tabcolsep}{-2pt}
\begin{adjustbox}{min width=\textwidth}
\begin{tabular}{c c c c c c c c c c c c c c c c c}
\hline\hline
Field & IFRS & \multicolumn{2}{c}{Radio coordinates} & \multicolumn{2}{c}{Combined positional} &      \multicolumn{2}{c}{Distance from claimed IR}       & Reliability &  $S_{1.4\,GHz}$  & $S_{3.6\,\mu m}$ & $S_{4.5\,\mu m}$ &      $S_{1.4\,GHz}/S_{3.6\,\mu m}$     &      $S_{1.4\,GHz}/S_{4.5\,\mu m}$     &$S_{4.5\,\mu m}/S_{3.6\,\mu m}$& Spectral &  Optical  \\
                                                                                                                                                                                                                                             
name  & ID   &      \multicolumn{2}{c}{(J2000)}      &  \multicolumn{2}{c}{accuracy (arcsec)}  & \multicolumn{2}{c}{3.6/4.5\,$\mu$m couterpart (arcsec)} &             &                         &                          &                          &$S_{1.4\,GHz}/(3 \times N_{3.6\,\mu m}$)&$S_{1.4\,GHz}/(3 \times N_{4.5\,\mu m})$&                                               &  index   & magnitude \\
                                                                                                                                                                                                                                                                                                                                                                                                    
      &      &    (hh:mm:ss)    &    (dd:mm:ss)      &   $\sigma_{RA}$   &   $\sigma_{Dec}$    &          $\Delta$RA        &           $\Delta$Dec      &     (\%)    &          (mJy)          &        ($\mu$Jy)         &       ($\mu$Jy)          &                                                       &                                                       &                                               &          &           \\
                                                                                                                                                                                                                                                                                                                                                                                                    
 (1)  & (2)  &       (3)        &      (4)           &         (5)       &         (6)         &             (7)            &             (8)            &     (9)     &          (10)           &           (11)           &           (12)           &                          (13)                         &                         (14)                          &                      (15)                     &   (16)   &    (17)   \\
\hline
 & ES0463  & 00:34:10.14 & $-$43:56:25.5 & 1.25 & 1.76 &      &      &    & 0.14 $\pm$ 0.04  & $<$ 1.29           & $<$ 1.71           & $>$ 109               & $>$ 82                & \dots & & \\*
 & ES0593  & 00:35:10.80 & $-$43:46:37.2 & 1.22 & 1.73 & 1.71 & 1.65 & 91 & 0.17 $\pm$ 0.05  & 2.98 $\pm$ 0.71  & 4.54 $\pm$ 0.52  & 57 $\pm$ 22         & 37 $\pm$ 12         & 1.52 $\pm$ 0.40 &                & \\*
 & ES0696  & 00:34:02.26 & $-$43:40:08.5 & 1.02 & 1.02 & 0.61 &      & 98 & 0.49 $\pm$ 0.11  & 1.33 $\pm$ 0.43  & $<$ 1.83           & 368 $\pm$ 145       & $>$ 268               & $<$ 1.38          &       0.07     & \\*
\multirow{3}{*}{\rotatebox{90}{ELAIS-S1}} & ES0913  & 00:37:33.42 & $-$43:24:53.4 & 1.02 & 1.02 & 0.63 &      & 98 & 0.68 $\pm$ 0.14  & 1.33 $\pm$ 0.39  & $<$ 1.47           & 511 $\pm$ 183       & $>$ 463               & $<$ 1.11          &     $-$1.62    & \\*
 & ES0973  & 00:38:44.13 & $-$43:19:20.4 & 1.32 & 1.44 &      &      &    & 9.14 $\pm$ 1.83  & $<$ 2.46           & $<$ 2.64           & $>$ 3715              & $>$ 3462              & \dots           &     $-$1.15    & \\*
 & ES1118  & 00:36:22.25 & $-$43:10:15.0 & 1.06 & 1.50 &      &      &    & 0.51 $\pm$ 0.11  & $<$ 1.41           & $<$ 1.56           & $>$ 362               & $>$ 327               & \dots           &                & \\*
 & ES1154  & 00:35:46.92 & $-$43:06:32.4 & 1.02 & 1.10 & 0.85 & 0.46 & 97 & 0.53 $\pm$ 0.11  & 2.11 $\pm$ 0.38  & 3.91 $\pm$ 0.53  & 251 $\pm$ 69        & 136 $\pm$ 34        & 1.85 $\pm$ 0.42 &     $-$0.56    & \\*
 & ES1193  & 00:37:19.58 & $-$43:02:01.4 & 1.07 & 1.52 &      &      &    & 0.23 $\pm$ 0.06  & $<$ 1.35           & $<$ 1.65           & $>$ 170               & $>$ 139               & \dots           &               & \\*
 & ES1259  & 00:38:27.17 & $-$42:51:33.7 & 1.02 & 1.02 &      &      &    & 4.52 $\pm$ 0.90  & $<$ 2.19           & out              & $>$ 2063              & \dots               & \dots           &               & \\*
 & ES1260  & 00:38:24.94 & $-$42:51:37.9 & 1.32 & 1.86 &      &      &    & 0.80 $\pm$ 0.16  & $<$ 1.92           & out              & $>$ 417               & \dots               & \dots           &               & \\* 
 &         &             &             &      &      &      &      &    &                  &                  &                  &                     &                     &                 &                & \\* 
 & DRAO3   & 16:05:30.48 & +54:09:02.0 & 0.26 & 0.26 & 0.82 & 1.14 & 97 & 3.56 $\pm$ 0.24  & 5.16 $\pm$ 0.43  & 6.38 $\pm$ 0.74  & 690 $\pm$ 74        & 558 $\pm$ 75        & 1.24 $\pm$ 0.18 &                & \\*
 & DRAO6   & 16:06:47.93 & +54:15:10.5 & 0.28 & 0.28 &  --  &  --  & -- & 2.23 $\pm$ 0.18  &        --        &        --        &        --           &        --           &        --       &       --       & -- \\*
\multirow{3}{*}{\rotatebox{90}{ELAIS-N1}} & DRAO7   & 16:08:38.74 & +54:27:51.8 & 0.31 & 0.31 & 2.15 & 2.20 & 82 & 1.39 $\pm$ 0.13  & 2.09 $\pm$ 0.50  & 2.32 $\pm$ 0.50  & 665 $\pm$ 171       & 599 $\pm$ 141       & \dots & & \\*
 &         &             &             &      &      &      &      &    &                  & $<$ 1.50           & $<$ 1.50           & $>$ 927               & $>$ 927               & \dots           &                & \\*
 & DRAO8   & 16:09:49.75 & +54:08:33.3 & 0.36 & 0.36 &      &      &    & 1.08 $\pm$ 0.13  & $<$ 1.47           & $<$ 1.17           & $>$ 735               & $>$ 923               & \dots           &                & \\*
 & DRAO9   & 16:11:12.89 & +54:33:17.6 & 0.25 & 0.25 &      &      &    & 2.45 $\pm$ 0.14  & $<$ 1.32           & $<$ 1.50           & $>$ 1856              & $>$ 1633              & \dots           &                & \\*
 & DRAO10  & 16:12:12.29 & +55:23:02.1 & 0.20 & 0.20 &  --  &  --  & -- & 360.15 $\pm$ 4.20&        --        &        --        &          --         &         --          &        --       &       --       & -- \\*
 & DRAO11  & 16:12:25.78 & +54:55:03.0 & 0.36 & 0.36 & 2.26 & 2.36 & 80 & 0.84 $\pm$ 0.10  & 7.88 $\pm$ 0.40  & 12.18 $\pm$ 0.66 & 107 $\pm$ 14        & 69 $\pm$ 9          & 1.55 $\pm$ 0.11 &                & \\*
 &         &             &             &      &      &      &      &    &                  & $<$ 1.20           & $<$ 1.98           & $>$ 700               & $>$ 424               & \dots           &                & \\*
 &         &             &             &      &      &      &      &    &                  &                  &                  &                     &                     &                 &                & \\*
 & LH5549   & 10:44:13.17 & +58:48:33.3 & 0.27 & 0.30 & 1.32 & 0.90 & 92 & 3.24 $\pm$ 0.13  & 3.01 $\pm$ 0.37  & 3.67 $\pm$ 0.58  & 1076 $\pm$ 139      & 883 $\pm$ 144      & 1.22 $\pm$ 0.24 &                & \\*
 & LH4124   & 10:44:13.77 & +58:17:45.3 & 0.75 & 0.90 & 1.48 & 1.94 & 89 & 0.50 $\pm$ 0.08  & 1.85 $\pm$ 0.39  & 2.32 $\pm$ 0.56  & 270 $\pm$ 72        & 216 $\pm$ 62       & 1.25 $\pm$ 0.40 &                & \\*
 &          &             &             &      &      &      &      &    &                  & $<$ 1.17           & $<$ 1.68           & $>$ 427               & $>$ 298              & \dots           &                & \\*
 & LH5709   & 10:44:35.75 & +58:53:10.0 & 0.21 & 0.21 & 0.87 & 0.49 & 96 & 8.98 $\pm$ 0.12  & 2.56 $\pm$ 0.63  & 2.14 $\pm$ 0.58  & 3508 $\pm$ 865      & 4196 $\pm$ 1139    & 0.84 $\pm$ 0.31 &                & \\*
 & LH5705   & 10:46:04.53 & +58:53:19.0 & 0.36 & 0.41 & 1.05 &      & 95 & 0.61 $\pm$ 0.04  & 1.44 $\pm$ 0.44  & $<$ 1.71           & 424 $\pm$ 132       & $>$ 357              & $<$ 1.19          &                & \\*
 & LH4270   & 10:47:11.26 & +58:21:46.7 & 0.22 & 0.23 &      &      &    & 0.52 $\pm$ 0.01  & $<$ 1.32           & $<$ 1.56           & $>$ 394               & $>$ 333              & \dots           &                & \\*
 & LH5995   & 10:47:34.56 & +59:07:01.2 & 0.34 & 0.39 &      &      &    & 1.47 $\pm$ 0.09  & $<$ 1.47           & $<$ 1.47           & $>$ 1000              & $>$ 1000             & \dots           &                & \\*
 & LH0912   & 10:47:59.36 & +57:17:39.8 & 0.20 & 0.20 & 0.25 & 0.52 & 99 & 1.39 $\pm$ 0.01  & 3.08 $\pm$ 0.49  & 6.01 $\pm$ 0.68  & 451 $\pm$ 72        & 231 $\pm$ 26       & 1.95 $\pm$ 0.38 &                & \\*
 & LH0324   & 10:48:06.21 & +57:03:00.6 & 0.25 & 0.27 & 1.07 & 1.14 & 95 & 0.61 $\pm$ 0.02  & 2.53 $\pm$ 0.49  & 5.94 $\pm$ 0.64  & 241 $\pm$ 47        & 103 $\pm$ 12       & 2.35 $\pm$ 0.52 &                & \\*
 & LH6025   & 10:49:04.39 & +59:09:13.5 & 0.27 & 0.29 &      &      &    & 1.28 $\pm$ 0.05  & $<$ 1.14           & $<$ 1.32           & $>$ 1123              & $>$ 970              & \dots           &                & \\* 
 & LH5512   & 10:49:23.18 & +58:48:51.3 & 0.20 & 0.20 &      &      &    & 2.66 $\pm$ 0.01  & $<$ 1.23           & $<$ 1.50           & $>$ 2163              & $>$ 1773             & \dots           &                & \\* 
\rotatebox{90}{LH} & LH3817   & 10:49:48.97 & +58:12:19.6 & 0.20 & 0.20 & 0.27 & 1.04 & 99 & 1.53 $\pm$ 0.01  & 3.48 $\pm$ 0.44  & 4.06 $\pm$ 0.60  & 440 $\pm$ 56        & 377 $\pm$ 56        & 1.17 $\pm$ 0.23 & & 20.75$_K$ $^{(d)}$\\*
 & LH0576   & 10:49:56.77 & +57:10:41.2 & 0.21 & 0.22 &      &      &    & 1.15 $\pm$ 0.02  & $<$ 1.20           & $<$ 1.35           & $>$ 958               & $>$ 852              & \dots           &                & \\* 
 & LH0502   & 10:51:22.09 & +57:08:55.0 & 0.20 & 0.20 & 0.09 & 0.26 & 99 & 10.59 $\pm$ 0.02 & 4.83 $\pm$ 0.54  & 8.41 $\pm$ 0.58  & 2193 $\pm$ 245      & 1259 $\pm$ 87      & 1.74 $\pm$ 0.23 &                & 24.84$_R$ $^{(e)}$\\*
 & LH2316   & 10:51:30.82 & +57:44:08.0 & 0.21 & 0.21 &      &      &    & 0.97 $\pm$ 0.01  & $<$ 1.77           & $<$ 1.68           & $>$ 548               & $>$ 577              & \dots           &                & \\*
 & LH2633   & 10:51:38.11 & +57:49:56.8 & 0.20 & 0.20 & 0.10 & 0.12 & 99 & 2.54 $\pm$ 0.01  & 5.43 $\pm$ 0.33  & 10.00 $\pm$ 0.56 & 468 $\pm$ 28        & 254 $\pm$ 14       & 1.84 $\pm$ 0.15 &                & \\* 
 & LH0209   & 10:52:39.55 & +56:58:25.6 & 0.20 & 0.20 & 1.00 & 0.68 & 95 & 2.83 $\pm$ 0.02  & 3.61 $\pm$ 0.50  & 5.37 $\pm$ 0.67  & 784 $\pm$ 109       & 527 $\pm$ 66       & 1.49 $\pm$ 0.28 &                & 24.41$_R$ $^{(e)}$\\*
 &          &             &             &      &      &      &      &    &                  &                  &                  &                     &                    &                 &                & 24.38$_I$ $^{(e)}$ \\*
 & LH5785   & 10:53:18.14 & +58:56:22.8 & 0.20 & 0.20 & 0.80 &      & 97 & 78.43 $\pm$ 0.03 & 4.40 $\pm$ 0.48  & out              & 17825 $\pm$ 1945    & \dots              & \dots           &                & \\* 
 & LH2417   & 10:54:14.89 & +57:45:57.7 & 0.23 & 0.25 &      & 1.20 & 93 & 0.37 $\pm$ 0.01  & $<$ 1.38           & 2.27 $\pm$ 0.43  & $>$ 268               & 163 $\pm$ 47       & $>$ 1.64          &                & \\* 
 & LH4721   & 10:54:57.39 & +58:31:53.5 & 0.20 & 0.20 &      &      &    & 11.89 $\pm$ 0.01 & $<$ 2.19           & out              & $>$ 5429              & \dots              & \dots           &                & \\* 
 & LH0943   & 10:55:48.54 & +57:18:27.8 & 0.20 & 0.20 &      &      &    & 13.65 $\pm$ 0.01 & $<$ 1.32           & $<$ 1.53           & $>$ 10341             & $>$ 8922             & \dots           &                & \\* 
 & LH1019   & 10:55:56.59 & +57:19:59.6 & 0.21 & 0.21 &      &      &    & 0.80 $\pm$ 0.01  & $<$ 1.05           & $<$ 1.53           & $>$ 762               & $>$ 523              & \dots           &                & \\
\hline
\end{tabular}
\end{adjustbox}
\tablefoot{
\small{
\small{$^{(a)}$ \citet{Arn2001}, from Deep Public Survey \citep{Mig2007}; $^{(b)}$ \citet{Raf2011}, from MUSYC \citep{Gaw2006}; $^{(c)}$ \citet{Mig2007}, from Deep Public Survey \citep{Mig2007}; $^{(d)}$ \citet{Law2007}, from UKIDSS \citep{Law2007}; $^{(e)}$ \citet{Fot2012}, from IfA Deep Survey \citep{Bar2004}.}
}
}
\end{table}
\end{landscape}

\clearpage

\appendix
\section{Reliable counterparts of IFRSs} \label{App:IFRS_CP}

\begin{figure}[!h]
\centering
\begin{minipage}[!t]{1.00\textwidth}
\caption{ \label{fig:IFRS_images1} Images of the reliable (i.e.\ reliability $>$90\%) counterparts of IFRSs identified in CDFS, ELAIS-S1, and ELAIS-N1 fields. Also reported are the counterparts for the sources ES0436 and ES0593, which are not IFRSs as their $R_{3.6}$ are around 57. All the images are \mbox{$\sim$ 40$\arcsec$ $\times$ 40$\arcsec$} wide, North is up and East on the left. The (inverted colours) black-and-white background images are SERVS cutouts, taken from the 3.6 or the 4.5\,$\mu$m mosaics (depending on which band the reliability has been computed on). Superimposed are a red cross (marking the radio peak position, always at the centre of the cutout), and a cyan circle (marking the area within which aperture photometry was derived).}
\end{minipage}
\begin{tabular}{ *3{>{\centering\arraybackslash}p{0.3\textwidth}} }
\mbox{}                                                                     & \mbox{}                                                                     & \mbox{}                                                                     \\
\bf{CS0114}                                                                 & \bf{CS0164}                                                                 & \bf{CS0173}                                                                 \\
\subfloat{\includegraphics[width=140pt]{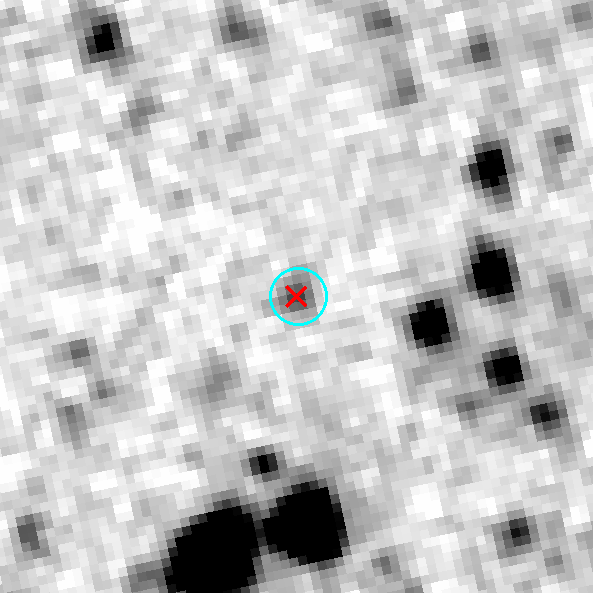}} & \subfloat{\includegraphics[width=140pt]{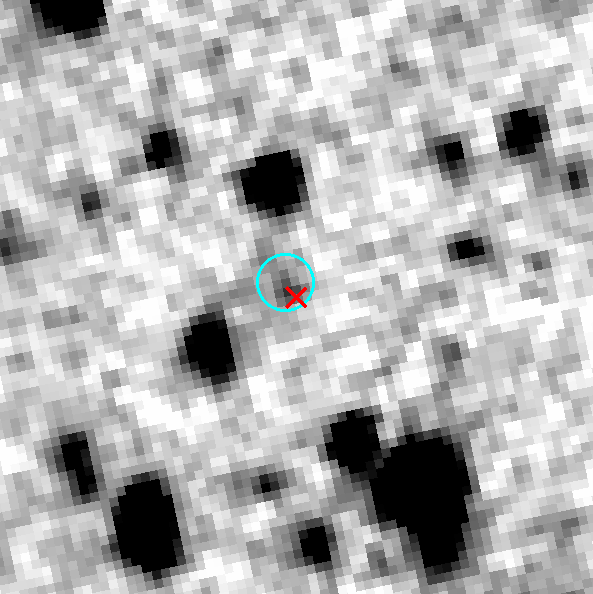}} & \subfloat{\includegraphics[width=140pt]{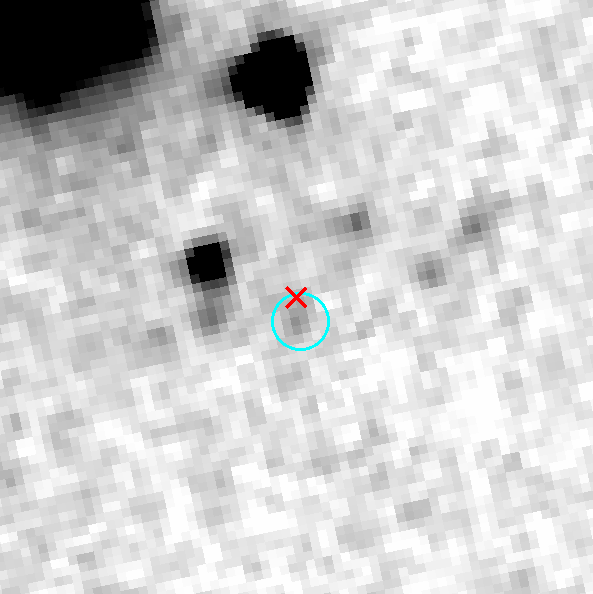}} \\
\mbox{}                                                                     & \mbox{}                                                                     & \mbox{}                                                                     \\
\mbox{}                                                                     & \mbox{}                                                                     & \mbox{}                                                                     \\
\bf{CS0255}                                                                 & \bf{CS0682}                                                                 & \bf{ES0056}                                                                 \\
\subfloat{\includegraphics[width=140pt]{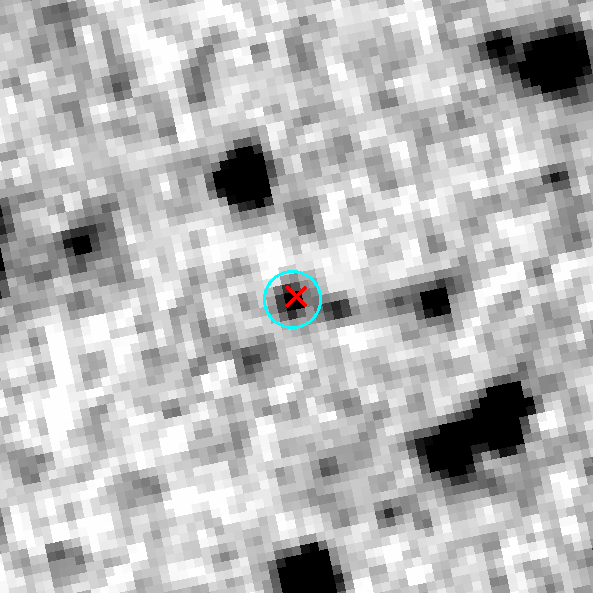}} & \subfloat{\includegraphics[width=140pt]{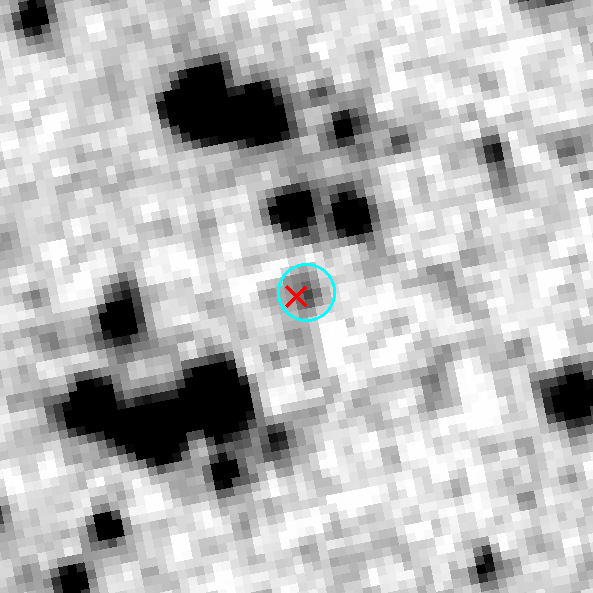}} & \subfloat{\includegraphics[width=140pt]{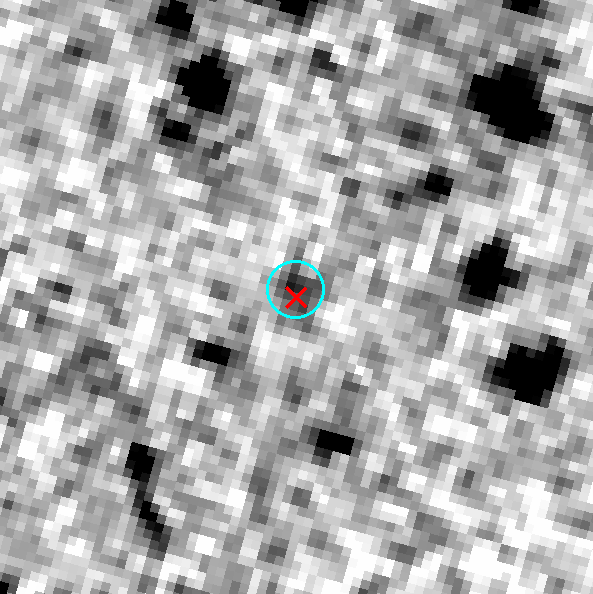}} \\
\mbox{}                                                                     & \mbox{}                                                                     & \mbox{}                                                                     \\
\mbox{}                                                                     & \mbox{}                                                                     & \mbox{}                                                                     \\
\bf{ES0156}                                                                 & \bf{ES0427}                                                                 & \bf{ES0436}                                                                 \\
\subfloat{\includegraphics[width=140pt]{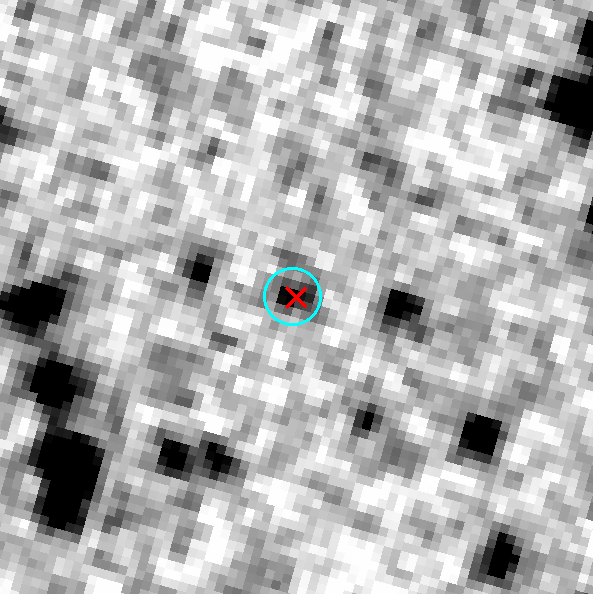}} & \subfloat{\includegraphics[width=140pt]{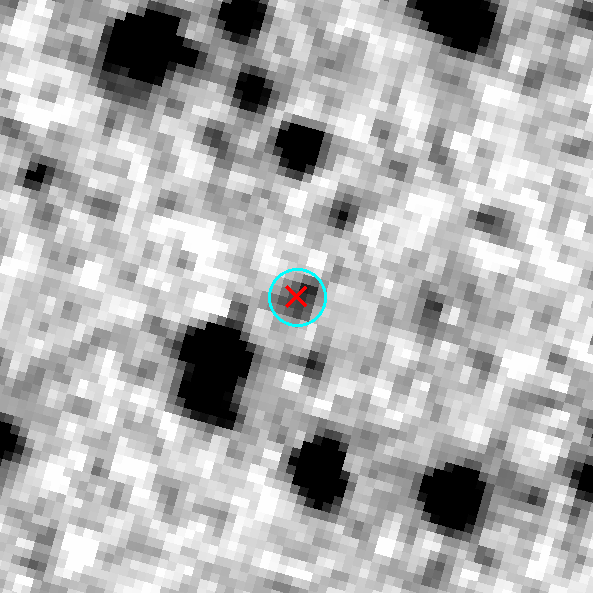}} & \subfloat{\includegraphics[width=140pt]{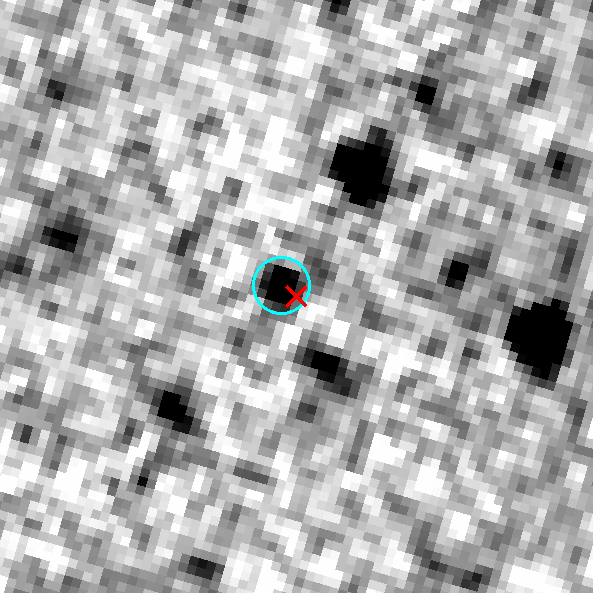}} \\
\end{tabular}
\end{figure}

\clearpage

\begin{figure}[!h]
\centering
\begin{tabular}{ *3{>{\centering\arraybackslash}p{0.3\textwidth}} }
\mbox{}                                                                     & \mbox{}                                                                     & \mbox{}                                                                    \\
\bf{ES0593}                                                                 & \bf{ES0696}                                                                 & \bf{ES0913}                                                                \\
\subfloat{\includegraphics[width=140pt]{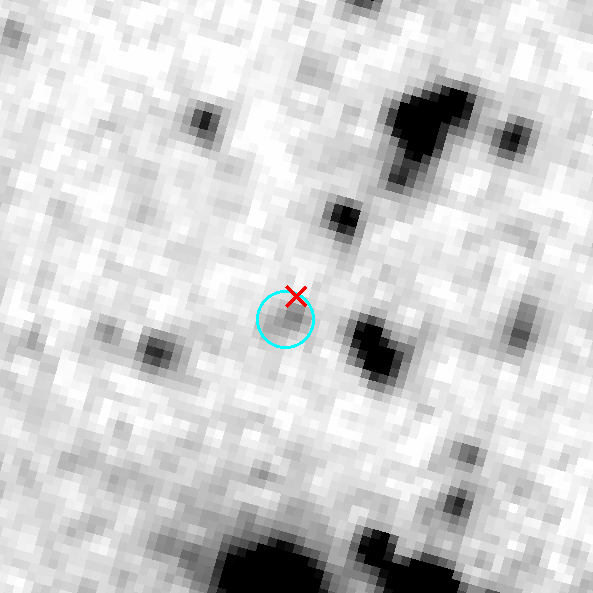}} & \subfloat{\includegraphics[width=140pt]{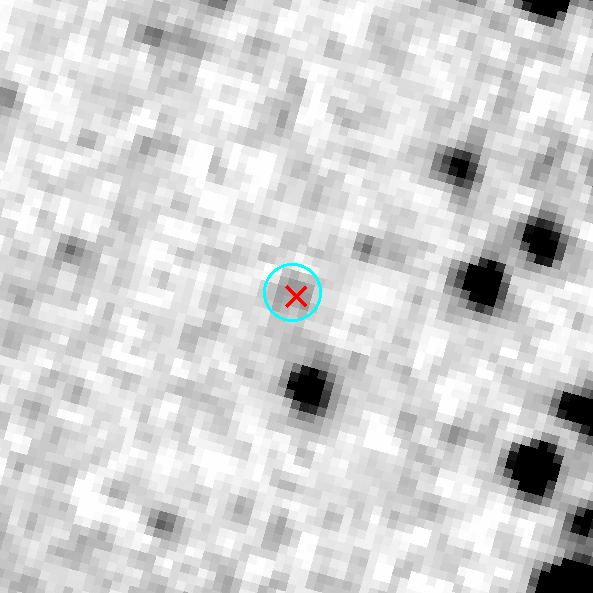}} & \subfloat{\includegraphics[width=140pt]{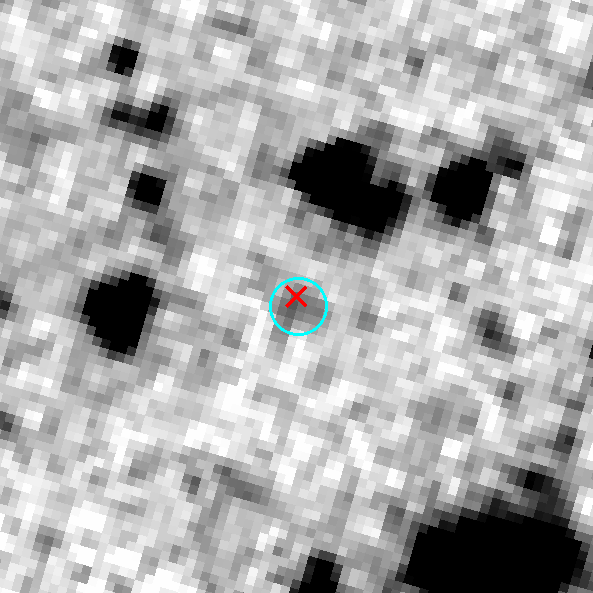}}\\
\mbox{}                                                                     & \mbox{}                                                                     & \mbox{}                                                                    \\
\mbox{}                                                                     & \mbox{}                                                                     & \mbox{}                                                                    \\
\bf{ES1154}                                                                 & \bf{DRAO3}                                                                  & \mbox{}                                                                    \\
\subfloat{\includegraphics[width=140pt]{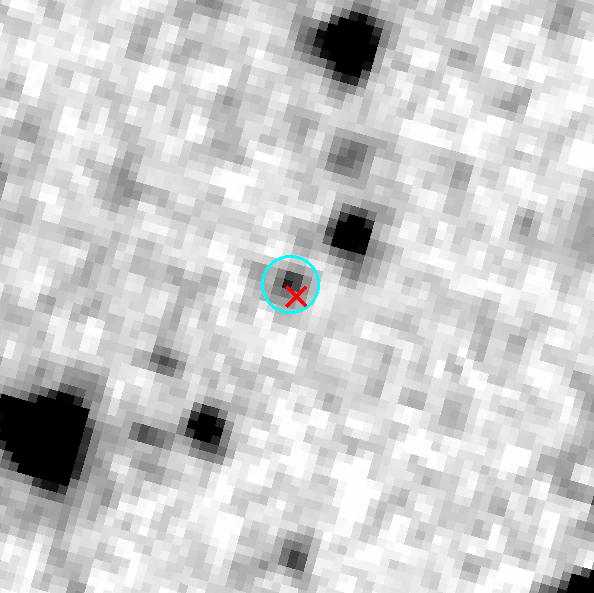}} & \subfloat{\includegraphics[width=140pt]{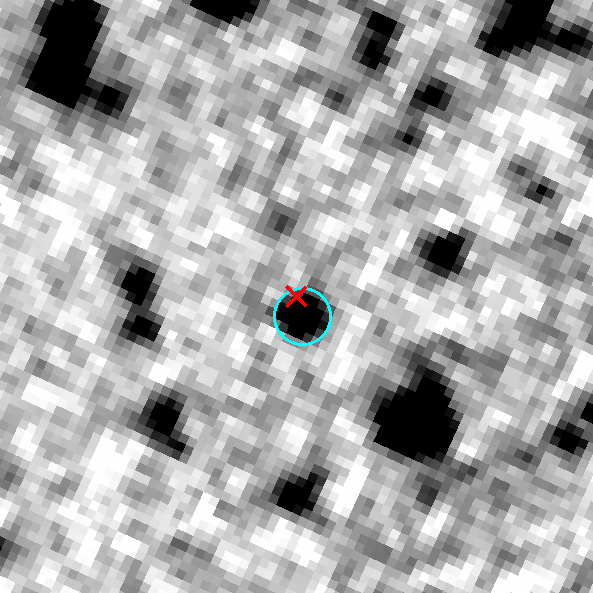}}  & \mbox{}                                                                    \\
\end{tabular}
\end{figure}

\clearpage

\begin{figure}[!h]
\centering
\begin{minipage}[!t]{1.00\textwidth}
\caption{ \label{fig:IFRS_images2} Images of the reliable (i.e.\ reliability $>$90\%) counterparts of IFRSs identified in LH SERVS field. All the images are \mbox{$\sim$ 40$\arcsec$ $\times$ 40$\arcsec$} wide, North is up and East on the left. The (inverted colours) black-and-white background images are SERVS cutouts, taken from the 3.6 or the 4.5\,$\mu$m mosaics (depending on which band the reliability has been computed). Superimposed are a red cross (marking the radio peak position, always at the centre of the cutout), a cyan circle (marking the area within which aperture photometry was derived), and a green square (marking the position of the optical counterparts, see Table \ref{tab:IFRS}). Also reported are the isophotes of the radio sources (magenta lines, contour levels at 10, 20, 40, 80, 160, 320, 640, and 1280$\sigma$ the radio image noise).}
\end{minipage}
\begin{tabular}{ *3{>{\centering\arraybackslash}p{0.3\textwidth}} }
\mbox{}                                                                  & \mbox{}                                                                  & \mbox{}                                                                  \\
\bf{LH5549}                                                              & \bf{LH5709}                                                              & \bf{LH5705}                                                              \\
\subfloat{\includegraphics[width=140pt]{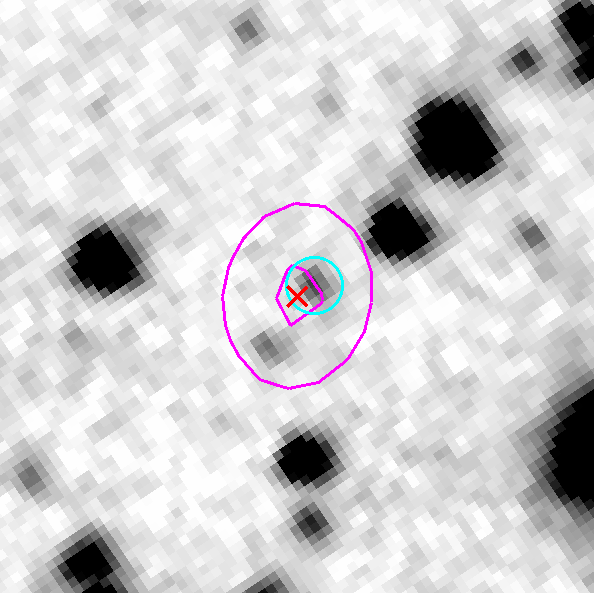}} & \subfloat{\includegraphics[width=140pt]{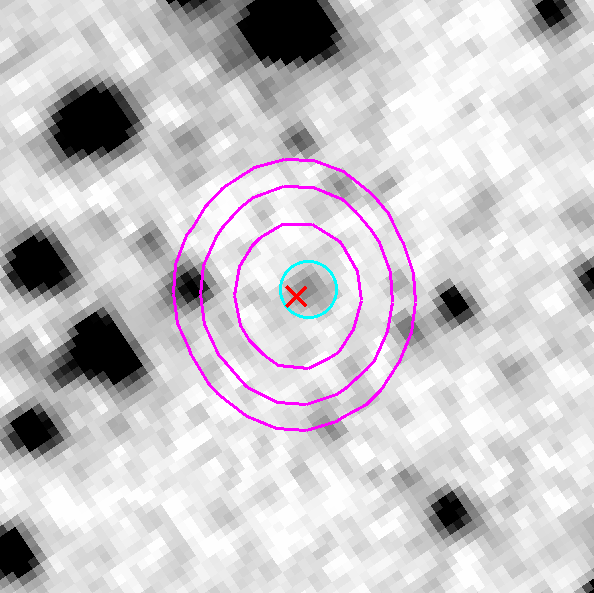}} & \subfloat{\includegraphics[width=140pt]{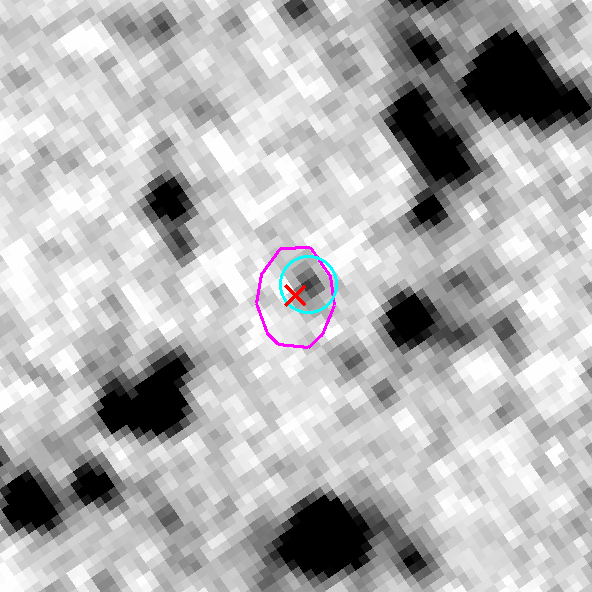}} \\
\mbox{}                                                                  & \mbox{}                                                                  & \mbox{}                                                                  \\
\mbox{}                                                                  & \mbox{}                                                                  & \mbox{}                                                                  \\
\bf{LH0912}                                                              & \bf{LH0324}                                                              & \bf{LH3817}                                                              \\
\subfloat{\includegraphics[width=140pt]{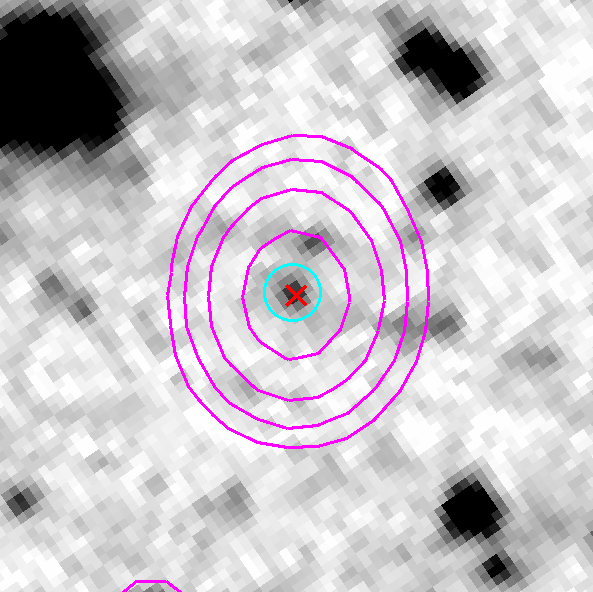}} & \subfloat{\includegraphics[width=140pt]{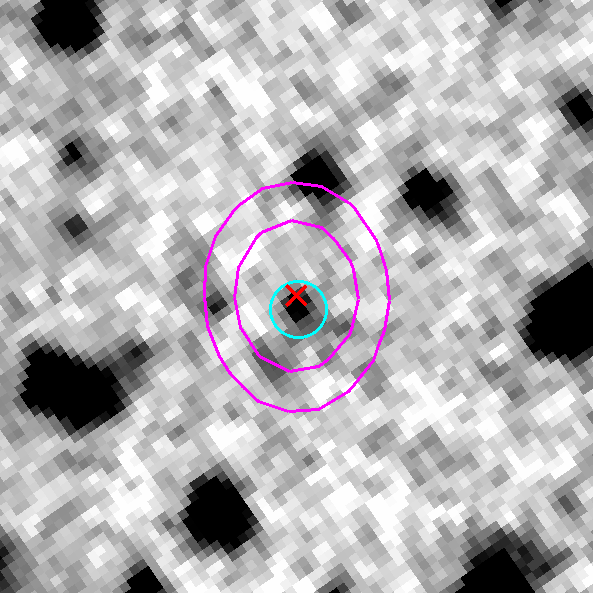}} & \subfloat{\includegraphics[width=140pt]{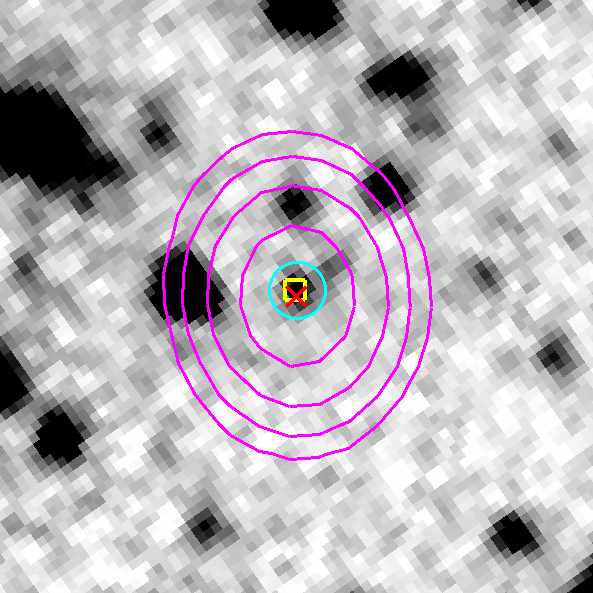}} \\
\mbox{}                                                                  & \mbox{}                                                                  & \mbox{}                                                                  \\
\mbox{}                                                                  & \mbox{}                                                                  & \mbox{}                                                                  \\
\bf{LH0502}                                                              & \bf{LH2633}                                                              & \bf{LH0209}                                                              \\
\subfloat{\includegraphics[width=140pt]{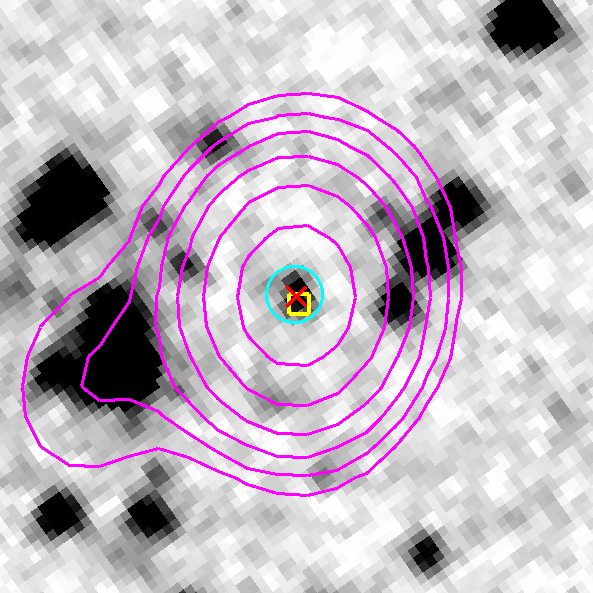}} & \subfloat{\includegraphics[width=140pt]{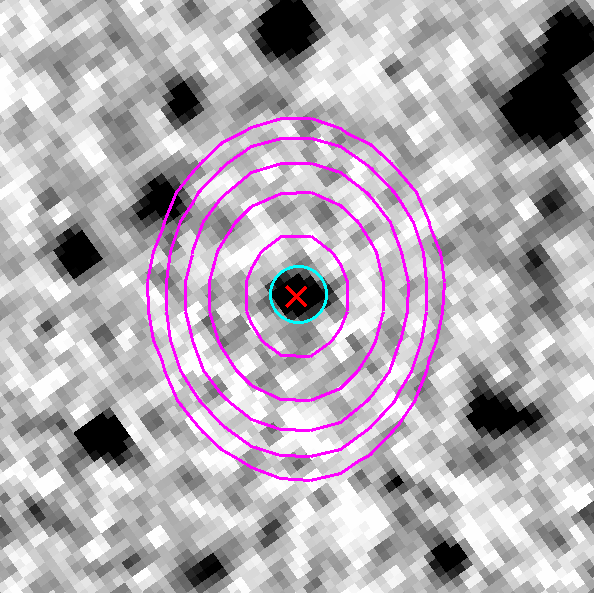}} & \subfloat{\includegraphics[width=140pt]{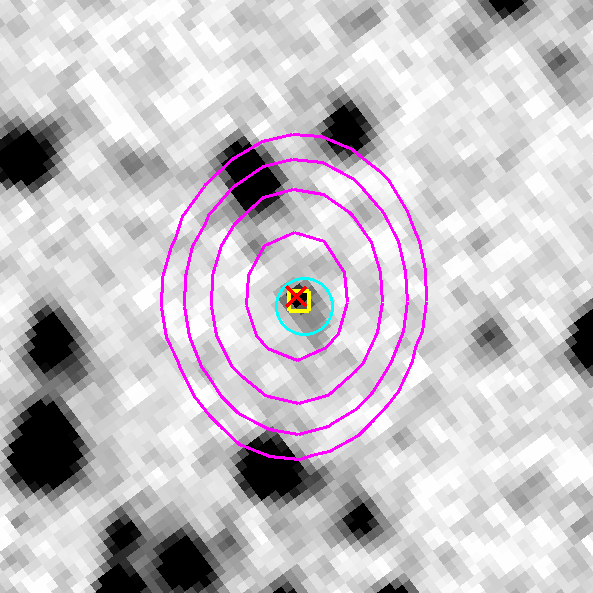}} \\
\end{tabular}
\end{figure}

\clearpage

\begin{figure}[!h]
\centering
\begin{tabular}{ *3{>{\centering\arraybackslash}p{0.3\textwidth}} }
\mbox{}                                                                  & \mbox{}                                                                  & \mbox{} \\
\bf{LH5785}                                                              & \bf{LH2417}                                                              & \mbox{} \\
\subfloat{\includegraphics[width=140pt]{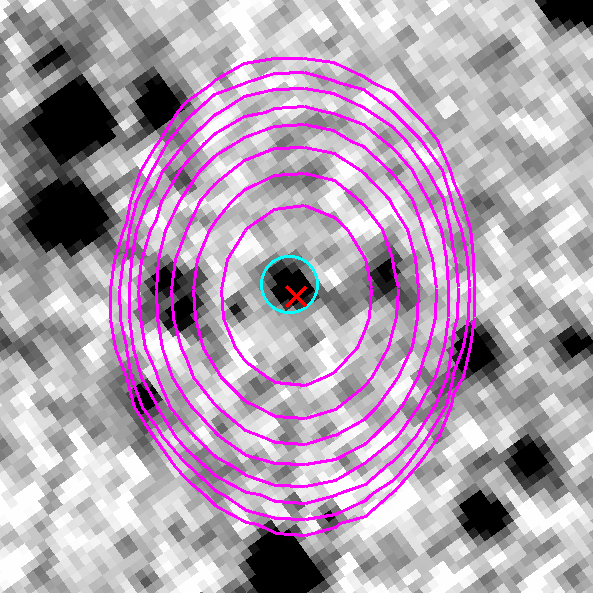}} & \subfloat{\includegraphics[width=140pt]{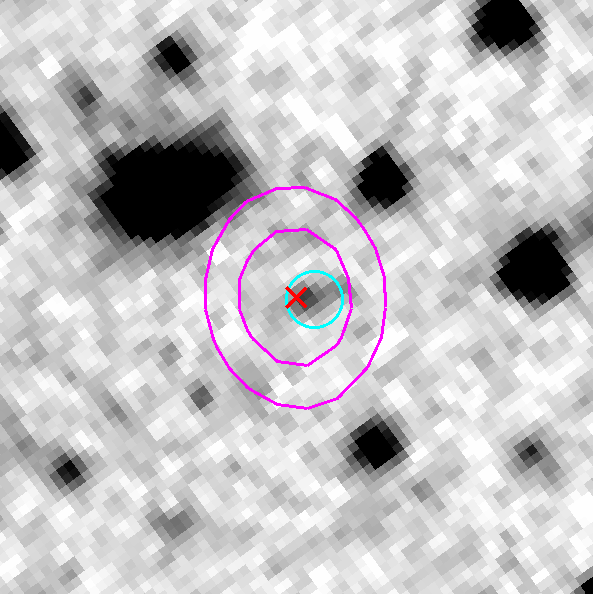}} & \mbox{} \\
\end{tabular}
\end{figure}

\end{document}